\documentclass{ws-mpla-FRKapr2025}  
\usepackage[super]{cite}
\usepackage{graphicx}

\newcommand{\beq}{\begin{equation}}
\newcommand{\eeq}{\end{equation}}
\newcommand{\beqa}{\begin{eqnarray}}
\newcommand{\eeqa}{\end{eqnarray}}
\newcommand{\bsubeqs}{\begin{subequations}}
\newcommand{\esubeqs}{\end{subequations}}

\makeatletter
\@addtoreset{equation}{section}
\renewcommand{\theequation}{\thesection.\arabic{equation}}
\makeatother

\begin{document}

\markboth{F.R. Klinkhamer}{Big Bang as spacetime defect}

%%%%%%%%%%%%%%%%%%%%% Publisher's Area please ignore %%%%%%%%%%%%%%
\catchline{}{}{}{}{}
%%%%%%%%%%%%%%%%%%%%%%%%%%%%%%%%%%%%%%%%%%%%%%%%%%%%%%%%%%%%%%%%%%%

\title{Big Bang as spacetime defect} 

\author{F.R. Klinkhamer}

\address{Institute for Theoretical Physics,
Karlsruhe Institute of Technology (KIT),\\
76128 Karlsruhe, Germany\\
frans.klinkhamer@kit.edu}

\maketitle

%%\pub{Received (Day Month Year)}{Revised (Day Month Year)}  

\begin{abstract}
We review the suggestion
that it is possible to eliminate the Big Bang curvature
singularity of the Friedmann cosmological solution by considering
a particular type of degenerate spacetime metric.
Specifically, we take the four-dimensional  
spacetime metric to have
a spacelike three-dimensional 
defect with a vanishing determinant
of the metric. This new solution suggests
the existence of another ``side'' of the Big Bang
(perhaps a more appropriate description
than ``pre-Big-Bang'' phase used in our original paper).
The corresponding new solution for defect wormholes is also
briefly discussed.
\end{abstract}
\vspace*{0.25\baselineskip}
{\footnotesize
\vspace*{0.75\baselineskip}
\noindent \hspace*{5mm}
\emph{Journal}: 
%%Mod. Phys. Lett. A in press 
Mod. Phys. Lett. A \textbf{40} (2025) 2530010
\vspace*{1.00\baselineskip}
\newline
\hspace*{5mm}
\emph{Preprint}: arXiv:2412.03538 %% (\version)  %%KA--TP--20--2024  
}
\vspace*{-5mm}\newline
\keywords{General relativity; big bang theory; spacetime topology.}
\ccode{PACS Nos.: 04.20.Cv, 98.80.Bp, 04.20.Gz.}

\section{Introduction}
\label{sec:Intro}

One of the great puzzles of modern cosmology, perhaps the greatest puzzle,
is the physical nature of the so-called Big Bang.
Concretely, the question is if there is really
an infinite energy density and an infinite spacetime
curvature or ``merely'' very large values for the energy density and
the spacetime curvature.

Modern cosmology relies
on the Hubble redshift--distance relation~\cite{Hubble1929}, which
is interpreted as an  \emph{expanding} Universe, where the adjective
refers to dropping values of the energy density of matter
as cosmic time advances.
The Big Bang is observed indirectly through
the Cosmic Microwave Background Radiation (CMBR)~\cite{PenziasWilson1965},
understood as the \emph{afterglow} of
a very hot and dense phase with temperatures $\gtrsim 4000\,\text{K}$,
now cooled down to a temperature of approximately
$3\,\text{K}$ by the adiabatic expansion of the Universe.
The observed photons of the CMBR were last scattered when the Universe
had a temperature of about $4000\,\text{K}$.
This cosmic Last Scattering Surface (LSS)
is directly analogous to the photosphere of the Sun, where the
photons are released that travel freely to the Earth.
These solar photons carry energy released by
nucleosynthesis in the dense and hot
environment at the center of the Sun.
Equally, the cosmic photons of the CMBR would have been
created in the dense and hot environment of the early Universe.

An even more indirect observation of the Big Bang is by
the measured nonzero abundances
of certain light elements, primarily helium $^{4}\text{He}$
but also traces of deuterium $^{2}\text{D}$,
the helium isotope $^{3}\text{He}$, and lithium $^{7}\text{Li}$.
As first realized by Gamov~\cite{Gamov1946},
these elements have been synthesized in an
even hotter and denser phase than the environment of
the LSS of the CMBR.
Typical temperatures of the primordial nucleosynthesis
are in a wide range
around $10^{9}\,\text{K}\sim 10^{5}\,\text{eV}$.

The implication of the CMBR and helium-abundance  measurements
is that there must have been a
very hot and dense epoch in the early history of our Universe.

The theoretical foundation of modern cosmology is provided by
Einstein's theory of
gravitation, known as the General Theory of Relativity
(often abbreviated as general relativity or GR)~\cite{Einstein1916}.
A particular cosmological solution of the corresponding
gravitational field equation
is given by the Friedmann solution~\cite{Friedmann1922},
with further input by Lema\^{i}tre~\cite{Lemaitre1931},
who considered the earliest phase in the history
of the Universe to correspond to
 ``un atome primitif''
(which has later been translated as ``a primeval atom'').
The mathematics of the relevant spacetime manifold was clarified
by Robertson~\cite{Robertson1935} and Walker~\cite{Walker1937}.
The Friedmann--Lema\^{i}tre--Robertson--Walker (FLRW)
cosmological solution, with a current epoch of Hubble expansion,
indicates the existence of an early moment
when the energy density and curvature were infinitely large;
this early moment was termed a ``big bang'' by Hoyle
in a 1949 BBC radio lecture~\cite{Peebles2009}.
As mentioned before, the question is whether or not this infinite curvature is a mathematical artifact

%%%%%%%%%%%%%%%%%\newpage%%tmp
Perhaps the suggested divergent values are real,
and we physicists have to completely revise of our understanding
of Nature. Indeed, we would need a new way to deal with these infinities
(not regularizing them as we are used to do in
elementary particle physics), while keeping the theory
under control and physically correct.

Or perhaps, more conservatively, the FLRW solution
is to be changed, so that the maximal energy density and
curvature values are very large but finite
(the typical energy scale may be the
so-called Planck scale~\cite{Planck1900}  given by
a combination of fundamental constants,
$E_\text{Planck} \equiv \sqrt{\hbar\,c^5/G}
\approx 1.22 \times 10^{19}\,\text{GeV}
\approx 1.42 \times 10^{32}\,\text{K}$).
It is clear that, in order to find such a new solution,
something needs to be changed. This change could,
for example, be an extended (or more flexible)
\emph{interpretation} of standard GR
or an entirely \emph{new} gravitational theory
(e.g., superstring theory, for which some references will
be given later).

In the present review, we follow the less radical approach of
keeping Einstein's gravitational field equation,
but allowing for different metrics than usually considered
and using a particular continuation procedure.
Specifically, we consider metrics with a vanishing determinant on a
measure-zero set of spacetime points,
namely a spacelike three-dimensional submanifold  
of the complete four-dimensional spacetime manifold.
This submanifold is interpreted as a ``spacetime defect,''
as will be explained later on.

The specific goal of this pedagogical review is threefold.
First, we try to give the simplest possible presentation of
our new defect solution
with a ``tamed'' Big Bang~\cite{Klinkhamer2019-prd}.
Second, we  make a few minor corrections and refinements.
Third, and most importantly,
we clarify the relevant mathematics of the new
solution, relying on earlier work by Horowitz~\cite{Horowitz1991}.

The outline of this review is then as follows. We first recall,
in Secs.~\ref{sec:Basics} and ~\ref{sec:FLRW cosmological solution},
the well-known equations of Friedmann cosmology
(mainly in order to establish notation)
and it is perfectly possible to skip
ahead to Sec.~\ref{sec:Defect cosmological solution}
for the new cosmological defect metric.
Having established what we may call ``defect cosmology''
(distinct from Friedmann cosmology), we next explore,
in Sec.~\ref{sec:Communication-between-the-two-worlds},
the important issue
of communication between the two ``sides''
of the Big Bang in defect cosmology
(in the usual bounce-cosmology interpretation,
between the ``pre-Big-Bang'' and ``post-big-bang'' phases).
Henceforth, we will call these two sides of the
Big-Bang  defect simply two ``worlds.''
As to the possible origin of this cosmological spacetime defect,
we present some brief remarks in
Sec.~\ref{sec:Nature-and-origin}.
We collect some final comments in Sec.~\ref{sec:Conclusion}.

There are also two appendices.
We, first, recall %% an ultraquick review of
the ``standard'' exotic-matter wormhole
in \ref{app:Recap exotic-matter-wormhole}
and, then, give
corresponding results for a new defect-wormhole solution in
\ref{app:Defect-wormhole solution},
again with special attention so some of the mathematical
subtleties (see, in particular,
\ref{subapp:Defect-WH-behavior-at-throat}).

%%\newpage%%tmp
\section{Basics}
\label{sec:Basics}

\subsection{Preliminary remarks}
\label{subsec:Preliminary remarks}

%%Time magazine, in its 31 December 1999 issue, named AE person of the century,

It is already over a century ago that Einstein realized that
the fabric of space and time gets deformed by the presence
of matter, and that space and time are not fixed once and
for all.  Elementary particles and their interaction fields are actors
on the stage of spacetime, and that stage responds to the activity of
the actors  (deformation of spacetime)
and reacts back on them (elasticity of spacetime).

Indeed, Einstein's insight was that spacetime is a dynamical
entity, responding to and interacting with the energy density
of ponderable matter.
Spacetime is then described by a Riemannian manifold with a metric
to define distances (strictly speaking, a pseudo-Riemannian manifold
as the square of the ``distance'' can be zero for certain pairs of
spacetime points and can even be negative for other pairs of points).
An alternative description is by tetrads (in a way, the square root
of the metric). Tetrads are essential to describe the propagation
of fermions (the Dirac equation being,
in a way, the square root of the Klein--Gordon equation).
Our main focus will be on the metric, but we will also mention
corresponding results for tetrads.

Natural units with $c=\hbar=k_{B}=1$ are used throughout.
The four-dimensional spacetime coordinate $x^{\mu}$ has
an index $\mu$ running over $\{0,\, 1,\, 2,\, 3 \}$
and the spacetime signature is \mbox{$(-+++)$.}
Occasionally, we also use
Lorentz indices $a$, $b$ running over $\{0,\, 1,\, 2,\, 3 \}$.

%%%%\newpage%%tmp
\subsection{Metric and Einstein field equation}
\label{subsec:Metric and Einstein field equation}

In this subsection, we recall the basic equations of GR and refer to
Weinberg's textbook~\cite{Weinberg1972} for notation and further references.
Also, we refer to the review~\cite{EguchiGilkeyHanson1980} for
some mathematical background.

The Einstein gravitational field equation can be obtained
by postulating an appropriate action:
\bsubeqs
\beqa
\label{eq:total-action}
S=S_G+S_{M}\,,
\eeqa
with the gravitational action
\beqa
\label{eq:G-action}
S_G= -\frac{1}{16 \pi G}\, \int \,d^4x\, \sqrt{-g}\;R\,,
\eeqa
and the matter action
\beqa
\label{eq:M-action}
S_{M}= \int \,d^4x\, \sqrt{-g}\;\mathcal{L}^\text{\,(M)}[\Phi]
\,,
\eeqa
\esubeqs
where $g$ is the determinant of the metric $g_{\mu\nu}$,
so that the combined volume element
$d^4x\, \sqrt{-g}$ is a scalar.
The gravitational action \eqref{eq:G-action}
takes the Einstein--Hilbert form with a single power of the
Ricci curvature scalar $R$ and
 $G$ is Newton's gravitational coupling constant.
The  matter action \eqref{eq:M-action} is the spacetime
integral of the matter Lagrange density $\mathcal{L}^\text{\,(M)}[\Phi]$,
here shown in terms of a single generic matter field $\Phi(x)$.

The gravitational field equation follows from
the vanishing variation of the action
\eqref{eq:total-action} with respect to the metric
[$g_{\mu\nu}(x) \to g_{\mu\nu}(x) + \delta g_{\mu\nu}(x)$]
and corresponds to the Einstein equation,
\begin{equation}
\label{eq:Einstein-eq}
G_{\mu\nu} \equiv R_{\mu\nu} - \frac12\, g_{\mu\nu}\,R =
- 8\pi G\;T_{\mu\nu}^\text{\,(M)} \,,
\end{equation}
where $G_{\mu\nu}$ is the Einstein tensor,   
$R_{\mu\nu}$ is 
the Ricci curvature tensor,
and $T_{\mu\nu}^\text{\,(M)}$ is 
the energy-mo\-men\-tum
tensor of the matter, which results from the variation of the matter
action \eqref{eq:M-action}:
\begin{equation}
\label{eq:T-munu-def}
T_{\mu\nu}^\text{\,(M)} =
\frac{2}{\sqrt{-g}}\;
\frac{\delta \mathcal{L}^\text{\,(M)}}{\delta g^{\mu\nu}}\,.
\end{equation}
The Einstein equation \eqref{eq:Einstein-eq} is a second-order
\emph{partial} differential equation (PDE)
for the metric field $g_{\mu\nu}(x)$, as $R_{\mu\nu}$
and $R$ on the left-hand side
are linear in the second derivative of the metric
and the matter term on the right-hand side  is assumed
to have no such second derivatives or higher ones.

The matter field equation follows from the vanishing variation of
the action \eqref{eq:total-action} with respect to the matter field
[$\Phi(x) \to \Phi(x) + \delta \Phi(x)$].
For a single noninteracting scalar field $\phi(x)$,
we obtain the well-known Klein--Gordon
equation, $(\Box - m^2)\,\phi =0$, now
with a covariantized d'Alembert operator $\Box$.

%%%%\newpage%%tmp
In order to prepare for cosmological applications later on, we
already give the energy-momentum tensor of a perfect fluid:
\beq\label{eq:Tmunu-perfect-fluid}
T_{\mu\nu}^\text{\,(M,\;perfect\;fluid)}=
\left(P_{M}+\rho_{M}\right)\,U_{\mu}\,U_{\nu}+P_{M}\;g_{\mu\nu}\,,
\eeq
with a normalized four-velocity $U^{\mu}$ of the
comoving fluid element and scalars $P_{M}$ and $\rho_{M}$,
corresponding to the pressure and the energy density measured
in a localized inertial frame comoving with the fluid.

Incidentally, even the case of a possible nonzero cosmological constant
$\Lambda$ is covered~\cite{Zeldovich1968} by considering a
homogeneous perfect fluid component (labeled by $V$ for vacuum)
with equation-of-state $P_{V}=-\rho_{V}$ and $\rho_{V}=\Lambda \geq 0$.
For the present discussion, we set $\Lambda=0$ and only consider
ponderable matter  (labeled by  $M$) .

%%%%\newpage%%tmp
\subsection{Tetrads, spin connection, and first-order vacuum equations}
\label{subsec:Tetrads, spin connection, and first-order vacuum equations}

The first-order formulation of general relativity
(also known as the Palatini
formulation \cite{Schrodinger1950,Ferraris-etal-1982,Wald1984})
uses the tetrad $e^{a}_{\mu}(x)$ instead of the metric $g_{\mu\nu}(x)$
and
the affine spin connection $\omega^{\phantom{z}a}_{\mu\phantom{z}b}(x)$
instead of the Christoffel symbol $\Gamma^{\lambda}_{\mu\nu}(x)$.

The tetrad builds the metric tensor by the following relation:
\beq
\label{eq:metric-from-tetrads}
g_{\mu\nu}(x)
\equiv e^{a}_{\phantom{z}\mu}(x)\,e^{b}_{\phantom{z}\nu}(x)\,\eta_{ab}\,,
\eeq
with the Minkowski metric $\eta_{ab}$ given explicitly by
\beq
\label{eq:Minkowski-eta-ab}
\eta_{ab} =
\Big[ \text{diag}
\big(  -1,\,  1,\, 1 \,,\,1 \big)
\Big]_{ab}\,.
\eeq
It is standard practice to call $\mu$, $\nu$ Einstein indices and
$a$, $b$ Lorentz indices. Also, the Einstein summation convention is
assumed to hold with an implicit summation of  
matching upper and lower indices, 
for example $a$ and $b$
in \eqref{eq:metric-from-tetrads} are summed
over $\{0,\, 1,\, 2,\, 3 \}$.

An elegant formulation is due to Cartan and uses
differential forms, where we follow
the notation of Ref.~\refcite{EguchiGilkeyHanson1980}.
The curvature 2-form, in terms of the
connection 1-form $\omega^{a}_{\phantom{z}b}
\equiv \omega^{\phantom{z}a}_{\mu\phantom{z}b}\,\text{d}x^{\mu}$,
is then given by
\beq
\label{eq:curvature-2-form}
R^{\,a}_{\,\phantom{z}b} \equiv \text{d}\omega^{\,a}_{\,\phantom{z}b}+
\omega^{a}_{\phantom{z}c}  \wedge \omega^{c}_{\phantom{z}b}\,,
\eeq
with exterior derivative $\text{d}$ and wedge product $\wedge$
(some crucial properties are $\text{d}^{2} \equiv \text{d}\,\text{d} =0$
and $\text{d}x \wedge \text{d}y =-  \text{d}y \wedge \text{d}x$).
Then, the first-order equations of general relativity
without matter are~\cite{Horowitz1991}%
\bsubeqs\label{eq:first-order-eqs}
\beqa\label{eq:first-order-eqs-no-torsion}
e^{\,[\,a} \wedge D\, e^{\,b\,]} &=& 0\,,
\\[2mm]
\label{eq:first-order-eqs-Ricci-flat}
e^{\,b} \wedge R^{\,cd}\,\epsilon_{abcd} &=& 0 \,,
\eeqa
\esubeqs
with the completely antisymmetric symbol $\epsilon_{abcd}$
and the square brackets around the Lorentz indices denoting
antisymmetrization. The covariant derivative appearing in
\eqref{eq:first-order-eqs-no-torsion} is defined by
\beq
\label{eq:covariant-derivative-1-form}
D\, e^{b} \equiv \text{d}e^{b}+\omega^{\,b}_{\,\phantom{z}c}\wedge e^{c}\,,
\eeq
which is analogous to the covariant derivative of
Yang--Mills theory in elementary particle physics.

The vacuum equations \eqref{eq:first-order-eqs}
for $e^{a}$ and $\omega^{a}_{\phantom{z}b}$
are manifestly first order, as
both $D$ and $R^{\,cd}\,$ carry a single exterior derivative $\text{d}$.
Equation \eqref{eq:first-order-eqs-no-torsion}
corresponds to the no-torsion condition
and \eqref{eq:first-order-eqs-Ricci-flat} to the Ricci-flatness equation
[$\mathcal{R}_{\mu\nu}(x) \equiv
\mathcal{R}^{\lambda}_{\,\phantom{z}\mu\lambda\nu}(x)=0$
in the standard coordinate formulation
with Ricci and Riemann tensors denoted by calligraphic symbols].

%%\newpage%%tmp
\section{FLRW cosmological solution}
\label{sec:FLRW cosmological solution}

\subsection{Robertson--Walker metric and Friedmann equations}
\label{subsec:RW metric and Friedmann equations}

Modern cosmology starts from
the Friedmann cosmological solution of the
Einstein gravitational equation for
matter given by a homogeneous perfect fluid.
The solution can be either expanding or contracting,
depending on the boundary conditions.
With the current epoch of Hubble expansion as boundary condition,
the relevant Friedmann solution describes an expanding universe.

The details of this cosmological solution are as follows.
For a homogeneous and isotropic cosmological model,
the relevant spatially-flat Robertson--Walker (RW)
metric is~\cite{Robertson1935,Walker1937}
\beq\label{eq:RW-ds2}
ds^{2}\,\Big|^\text{(RW)}
\equiv
g_{\mu\nu}(x)\, dx^{\mu}\,dx^{\nu} \,\Big|^\text{(RW)}
=
- d t^{2}+ a^{2}( t )\;\delta_{m n}\,dx^{m}\,dx^n\,,
\eeq
with cosmic time coordinate $x^{0}=c\, t=t$
and spatial indices $m$, $n$  running over $\{1,\, 2,\, 3 \}$.
Again, the Einstein summation convention holds with
repeated indices $\mu,\,\nu$ and $m,\, n$ summed over
their respective ranges.
The Kronecker symbol $\delta_{m n}$ equals 1 for $m=n$
and $0$ otherwise.

For a homogeneous perfect fluid
with energy density $\rho_{M}(t)$ and pressure $P_{M}(t)$,
the Einstein equation \eqref{eq:Einstein-eq}
with the RW metric \textit{Ansatz} \eqref{eq:RW-ds2}
and the energy-momentum tensor \eqref{eq:Tmunu-perfect-fluid}
gives the spatially-flat Friedmann equations:%
\bsubeqs\label{eq:Feqs}
\beqa
\label{eq:1stFeq}
\hspace*{-0mm}&&
\left( \frac{\dot{a}}{a}\right)^{2}
= \frac{8\pi G}{3}\,\rho_{M}\,,
\\[2mm]
\label{eq:2ndFeq}
\hspace*{-0mm}&&
\frac{\ddot{a}}{a}+
\frac{1}{2}\,\left( \frac{\dot{a}}{a}\right)^{2}
=
-4\pi G\,P_{M}\,,
\\[2mm]
\label{eq:rhoMprimeeq}
\hspace*{-0mm}&&
\dot{\rho}_{M}+ 3\;\frac{\dot{a}}{a}\;
\Big(\rho_{M}+P_{M}\Big) =0\,,
\\[2mm]
\label{eq:M-EOS}
\hspace*{-0mm}&&
P_{M} = P_{M} \big(\rho_{M}\big)\,,
\eeqa
\esubeqs
where differentiation with respect to cosmic time $t$
has been denoted by an overdot.
The first three equations are first-order and second-order
\emph{ordinary} differential equations (ODEs) for the
variables $a(t)$ and $\rho_{M}(t)$,
with the third corresponding to energy conservation.
The fourth equation stands for the equation-of-state (EOS)
relation between pressure and energy density of the perfect fluid.
Furthermore, matter is assumed to obey the standard energy conditions.
Specifically, the null energy condition (NEC)
of the perfect fluid \eqref{eq:Tmunu-perfect-fluid}
corresponds to the inequality $\rho_{M}+P_{M} \geq 0$.

%%%%%\newpage%%tmp
\subsection{RW tetrads and spin connection}
\label{subsec:RW tetrads and spin connection}

For later discussion, we also give the tetrad and
spin connection corresponding to the RW metric \eqref{eq:RW-ds2},
where we will use the differential forms mentioned
in Sec.~\ref{subsec:Tetrads, spin connection, and first-order vacuum equations}.

The following dual basis
$e^{a} \equiv e^{a}_{\phantom{z}\mu}\,\text{d}x^{\mu}$\,
can be chosen:
\bsubeqs\label{eq:RW-tetrad}
\beqa
\label{eq:RW-tetrad-a-is-0}
e^{0}\,\Big|^\text{(RW)}
&=& \text{d}t\,,
\\[2mm]
\label{eq:RW-tetrad-a-is-m}
e^{m}\,\Big|^\text{(RW)}
&=& + a(t) \,\text{d}x^{m}\,,
\\[2mm]
\label{eq:RW-tetrad-a-positive}
a(t)\,\Big|^\text{(RW)}  &>& 0\,,
\eeqa
\esubeqs
for a spatial index $m \in \{1,\, 2,\, 3 \}$.
Note that, more generally, there could be various $\pm$
signs on the right-hand side of \eqref{eq:RW-tetrad-a-is-m},
but  three plus signs reproduce, for the case $a(t)=1$,
the flat-spacetime tetrad
$e^{a}_{\phantom{z}\mu}=\delta^{a}_{\phantom{z}\mu}$.

The metricity condition $\omega_{ab}=-\omega_{ba}$
and the no-torsion condition
$\text{d}e^{a}+\omega^{\,a}_{\,\phantom{z}b}\wedge e^{b} =0$,
listed as Eqs.~(3.10) and (3.11) in Ref.~\refcite{EguchiGilkeyHanson1980},
determine the spin connection $\omega^{\,a}_{\,\phantom{z}b}(x)$,
which has the following nonzero components:
\beqa
\label{eq:RW-connection}
\omega^{m}_{\phantom{z}0}\,\Big|^\text{(RW)}
&=&
-\omega^{0}_{\phantom{z}m}\,\Big|^\text{(RW)}
=
\dot{a}\,\text{d}x^{m}\,.
\eeqa
For the case $a(t)=1$,
we have $\omega^{\,a}_{\,\phantom{z}b}(x)=0$, which
gives for the curvature 2-form \eqref{eq:curvature-2-form}
the result $R^{\,a}_{\,\phantom{z}b}(x)=0$ corresponding to
flat Minkowski spacetime.

%%%%%\newpage%%tmp
\subsection{FLRW solution: Big Bang curvature singularity}
\label{subsec:Big Bang curvature singularitys}

The Friedmann equations \eqref{eq:Feqs}
for relativistic matter with constant EOS parameter
$w_{M} \equiv P_{M}/\rho_{M} =1/3$
have the well-known Friedmann--Lema\^{i}tre--Robertson--Walker (FLRW)
solution~\cite{Friedmann1922,Lemaitre1931,Robertson1935,Walker1937}:%
\bsubeqs\label{eq:Friedmann-solution}
\beqa
\label{eq:Friedmann-asol}
a(t)\,\Big|^\text{(FLRW)}_{(w_{M}=1/3)} &=& \sqrt{t/t_{0}}\,,
\phantom{\rho_{M0}/a^{4}(t) =\rho_{M0}\; t_{0}^{2}/t^{2}}
\hspace*{-4mm}
\text{for}\;\;\;t>0\,,
\\[2mm]
\label{eq:Friedmann-rhoMsol}
\rho_{M}(t)\,\Big|^\text{(FLRW)}_{(w_{M}=1/3)} &=&
\rho_{M0}/a^{4}(t) =\rho_{M0}\; t_{0}^{2}/t^{2}\,,
\phantom{\sqrt{t/t_{0}}}
\hspace*{-4mm}
\text{for}\;\;\;t>0\,,
\eeqa
\esubeqs
with cosmic scale factor normalized to
$a(t_{0})=1$ at $t_{0}>0$ and $\rho_{M0}$ a positive constant
[in fact, $G \rho_{M0}$ turns out to be proportional
to $1/t_0^{2}\,$, according to \eqref{eq:1stFeq}].

\begin{figure}[t]%[h]
\begin{center}
\hspace*{0mm}      %%renamed FIG1-v1.eps=FIG01-v4.eps
\includegraphics[width=0.55\textwidth]{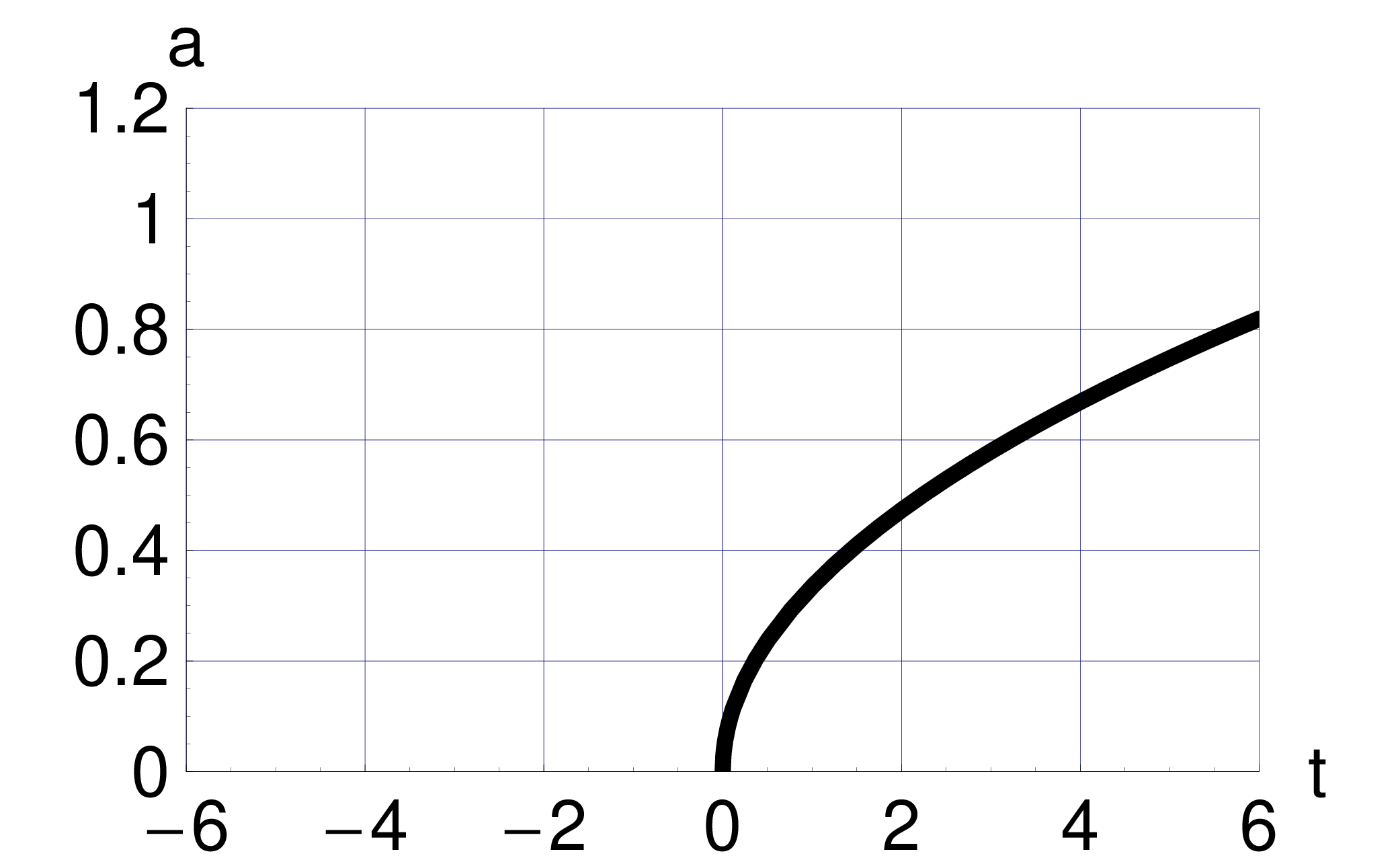}
%%{BB-as-spacetime-defect-FIG1-v012.eps}
\end{center}
\vspace*{8pt}
\caption{Cosmic scale factor $a(t)$ of the spatially-flat
FLRW universe with $w_{M}=1/3$ matter,
as given by \eqref{eq:Friedmann-asol} with $t_{0}=4\,\sqrt{5}$.
\vspace*{-0mm}}
\protect\label{fig:a-FLRW}
\end{figure}

The FLRW solution \eqref{eq:Friedmann-asol},
shown in Fig.~\ref{fig:a-FLRW},
displays the big bang singularity for $t\to 0^{+}$:%
\beq
\lim_{t\to 0^{+}} a(t) =0\,,
\eeq
with diverging curvature and energy density. Indeed, we have
a diverging Kretschmann curvature scalar
$K\equiv R^{\mu\nu\rho\sigma}\,R_{\mu\nu\rho\sigma}$,
\bsubeqs
\beq
\label{eq:Friedmann-sol-K-div}
K\,\Big|^\text{(FLRW)}_{(w_{M}=1/3)}
\sim 1/t^{4} \to \infty \;\;\text{for}\;\;t\to 0^{+}\,,
\eeq
and a diverging energy density,
\beq
\label{eq:Friedmann-sol-rhoM-div}
\rho_{M}(t)\,\Big|^\text{(FLRW)}_{(w_{M}=1/3)}
\sim 1/t^{2} \to \infty \;\;\text{for}\;\;t\to 0^{+}\,.
\eeq
\esubeqs
The affine spin connection  \eqref{eq:RW-connection}
of the FLRW solution also diverges as $1/\sqrt{t}$
for $t\to 0^{+}$.

As explained in the Introduction,
the question is whether or not these infinities are
mathematical artifacts.
Incidentally, inflation~\cite{Guth1981,Linde1982,AlbrechtSteinhardt1982}
does not really solve the problem
of proper initial conditions
(see, for example, the discussion in
Secs.~27.13 and 28.5 of Ref.~\refcite{Penrose2005}).

%%\newpage%%tmp
\section{Defect cosmological solution}
\label{sec:Defect cosmological solution}

\subsection{Defect cosmological metric and modified Friedmann equations}
\label{subsec:Defect cosmological metric and modified Friedmann equations}

We start by replacing the original
RW metric~\cite{Robertson1935,Walker1937}
with the following \textit{Ansatz}~\cite{Klinkhamer2019-prd}:
\bsubeqs\label{eq:RWK}
\beqa\label{eq:RWK-ds2}
\hspace*{-4mm}
ds^{2}\,\Big|^\text{(RWK)}
&\equiv&
g_{\mu\nu}(x)\, dx^{\mu}\,dx^{\nu} \,
\Big|^\text{(RWK)}
=
- \frac{t^{2}}{t^{2}+b^{2}}\,d t^{2}
+ a^{2}( t )
\;\delta_{m n}\,dx^{m}\,dx^{n}\,,
\\[2mm]
\label{eq:RWK-b-positive}
\hspace*{-4mm}
b &>& 0\,,
\\[2mm]
\hspace*{-4mm}
a( t ) &>& 0\,,
\\[2mm]
\label{eq:RWK-ranges-coordinates}
\hspace*{-4mm}
 t    &\in& (-\infty,\,\infty)\,,\quad
 x^{m} \in (-\infty,\,\infty)\,,
\eeqa
\esubeqs
for $x^{0}=t$.
The length scale $b$ entering
the metric component $g_{00}(t)$ must be nonzero and
is taken to be positive by convention. Equally,
the factor $a^{2}( t )$ of the spatial metric components $g_{mm}(t)$
must be nonzero, and we take $a( t ) > 0$ for regular behavior
of the spin connection, as will be discussed in
Sec.~\ref{subsec:Defect cosmological tetrads and spin connection}.

Formally, setting $b=0$ in the metric \eqref{eq:RWK-ds2}
reproduces the RW metric \eqref{eq:RW-ds2} with $g_{00}=-1$,
but this does not really hold for the metric \eqref{eq:RWK}
with $b>0$ at $t=0$, which has $g_{00}(0)=0$.
Apparently, the limits $b \to 0$ and $t \to 0$
do not commute for $g_{00}(t)$ from \eqref{eq:RWK}.

Expanding on the last remark,
there is a direct relation between the
RW metric \eqref{eq:RW-ds2}
and the new metric \eqref{eq:RWK},
but with a subtlety.
Changing the time coordinate $t\in  (-\infty,\,\infty)$ to
a new time coordinate $\tau \in  (-\infty,\,-b]  \, \cup \, [b,\,\infty)$
by the following coordinate transformation:
\beqa
\label{eq:tau-from-t}
\tau&=&
\begin{cases}
 + \sqrt{b^{2}+t^{2}}\,,    & \;\;\text{for}\;\; t \geq 0\,,
 \\[2mm]
 - \sqrt{b^{2}+t^{2}}\,,    & \;\;\text{for}\;\; t \leq 0\,,
\end{cases}
\eeqa
turns the metric \eqref{eq:RWK} into the RW metric
\eqref{eq:RW-ds2} in terms of $\tau$:
\beq\label{eq:RW-ds2-for-tau-coord}
ds^{2} =
- d \tau^{2}+ a^{2}( \tau )\;\delta_{m n}\,dx^{m}\,dx^n\,.
\eeq
But the coordinate
transformation \eqref{eq:tau-from-t}
is \emph{multivalued} at $t=0$ (and the inverse transformation is
discontinuous).
Hence, the coordinate transformation
from $t$ to $\tau$ is \emph{not} a diffeomorphism
(an invertible $\text{C}^{\infty}$ function by definition).
In short, the differential structure
of the metric \eqref{eq:RWK-ds2} in terms of $t$
is different from that of the standard spatially-flat FLRW metric
\eqref{eq:RW-ds2-for-tau-coord}
in terms of $\tau$, as will be discussed below.

The heart of the matter is that the
metric $g_{\mu\nu}(x)$ from \eqref{eq:RWK}
is \emph{degenerate}, with a vanishing determinant at $t = 0$.
The $t = 0$ slice corresponds
to a three-dimensional \emph{spacetime defect}, where
the terminology emphasizes the analogy with a crystallographic
defect in an atomic crystal~\cite{Mermin1979}.
For some background on this type of spacetime defect, we refer to the
papers~\cite{Schwarz2010,Klinkhamer2014-mpla,Klinkhamer2014-prd,KlinkhamerSorba2014,%
Guenther2017}
and a subsequent review~\cite{Klinkhamer2019-JPCS}
(related papers~\protect\cite{Klinkhamer2020-more,KlinkhamerWang2019-cosm,%
KlinkhamerWang2020-pert,Wang2021,Battista2021,Klinkhamer-Epiphany2021}
for the cosmological defect will be discussed later).

%%%%\newpage%%tmp
Inserting the new metric \textit{Ansatz} \eqref{eq:RWK}
and the energy-momentum tensor \eqref{eq:Tmunu-perfect-fluid}
of a homogeneous perfect fluid
into the standard Einstein equation \eqref{eq:Einstein-eq}
gives \emph{modified} spatially-flat Friedmann equations:%
\bsubeqs\label{eq:mod-Feqs}
\beqa
\label{eq:mod-1stFeq}
\hspace*{-0mm}&&
\left[1+ \frac{b^{2}}{t^{2}}\,\right]\,
\left( \frac{\dot{a}}{a}\right)^{2}
= \frac{8\pi G}{3}\,\rho_{M}\,,
\\[2.00mm]
\label{eq:mod-2ndFeq}
\hspace*{-0mm}&&
\left[1+\frac{b^{2}}{t^{2}}\,\right]\,
\left(\frac{\ddot{a}}{a}+
\frac{1}{2}\,\left( \frac{\dot{a}}{a}\right)^{2}
\right)
-\frac{b^{2}}{t^{3}}\,\frac{\dot{a}}{a}
=
-4\pi G\,P_{M}\,,
\\[2.00mm]
\label{eq:mod-Feq-rhoMprimeeq}
\hspace*{-0mm}&&
\dot{\rho}_{M}+ 3\;\frac{\dot{a}}{a}\;
\Big(\rho_{M}+P_{M}\Big) =0\,,
\\[2.00mm]
\label{eq:mod-Feq-EOS}
\hspace*{-0mm}&&
P_{M} = P_{M} \big(\rho_{M}\big)\,,
\eeqa
\esubeqs
where the overdot stands again for differentiation with respect to $t$.
Two remarks are in order.
First, the inverse metric from \eqref{eq:RWK-ds2}
has a component $g^{00}=(t^{2}+b^{2})/t^{2}$
that diverges at $t=0$  and we must be careful to obtain
the reduced field equations at $t=0$ from the limit $t \to 0$;
see Sec.~\ref{subsec:Mathematics-Continuous-extension}
for further discussion.

Second, the new $b^{2}/t^{2}$ terms in
the modified Friedmann equations \eqref{eq:mod-1stFeq}
and \eqref{eq:mod-2ndFeq} are a manifestation of the
different differential structure of  \eqref{eq:RWK-ds2}
compared to the differential structure of \eqref{eq:RW-ds2}
which gives the standard Friedmann equations \eqref{eq:1stFeq}
and \eqref{eq:2ndFeq}.

Expanding on the last point, we note that,  %%xxx
away from the defect at $t = 0$ (or $\tau = \pm b$),
the \emph{local} physical effects
from the metric \eqref{eq:RWK-ds2}
are the same as those from the
standard RW metric \eqref{eq:RW-ds2-for-tau-coord},
but \emph{global} properties may be different due to different
boundary conditions at $t=0$. Something similar has been observed
for pure-space defects,
where the different differential structure affects the
global aspects (parity) of the solutions of the Klein--Gordon
equation~\cite{KlinkhamerSorba2014}.

%%%%\newpage%%tmp
\subsection{Defect cosmological tetrads and spin connection}
\label{subsec:Defect cosmological tetrads and spin connection}

Let us now discuss the tetrad and
spin connection corresponding to the RWK metric \eqref{eq:RWK-ds2},
again using differential forms.

The following dual basis
$e^{a} \equiv e^{a}_{\phantom{z}\mu}\,\text{d}x^{\mu}$\,
can be chosen:
\bsubeqs\label{eq:RWK-tetrad}
\beqa
\label{eq:RWK-tetrad-a-is-0}
e^{0}\,\Big|^\text{(RWK)}
&=&  \frac{t}{\sqrt{t^{2}+b^{2}}}\,\text{d}t\,,
\\[2mm]
\label{eq:RWK-tetrad-a-is-m}
e^{m}\,\Big|^\text{(RWK)}
&=&  a(t) \,\text{d}x^{m}\,,
\\[2mm]
\label{eq:RWK-tetrad-a-positive}
a(t)\,\Big|^\text{(RWK)}  &>& 0\,,
\eeqa
\esubeqs
for a spatial index $m \in \{1,\, 2,\, 3 \}$.
Formally, setting $b=0$ in these RWK tetrads for $t>0$
reproduces the RW tetrads of \eqref{eq:RWK-tetrad}.

The no-torsion condition
$\text{d}e^{a}+\omega^{\,a}_{\,\phantom{z}b}\wedge e^{b} =0$
and the metricity condition $\omega_{ab}=-\omega_{ba}$
determine the connection $\omega^{a}_{\phantom{z}b}(x)$, which
has the following components:
\bsubeqs\label{eq:RWK-connection}
\beqa
\label{eq:RWK-connection-m-0}
\omega^{m}_{\phantom{z}0}\,\Big|^\text{(RWK)}
&=&
-\omega^{0}_{\phantom{z}m}\,\Big|^\text{(RWK)}
=
\frac{\sqrt{t^{2}+b^{2}}}{t}\;\dot{a}\;\text{d}x^{m}\,,
\\[2mm]
\omega^{0}_{\phantom{z}0}\,\Big|^\text{(RWK)}
&=&
0\,,
\\[2mm]
\omega^{m}_{\phantom{z}n}\,\Big|^\text{(RWK)}
&=&
0\,.
\eeqa
\esubeqs
For generic $a(t)$, the connection components \eqref{eq:RWK-connection-m-0}
diverge at $t=0$. But for a bounce-type behavior of $a(t)$ at $t=0$,
\bsubeqs\label{eq:bounce-type-a}
\beqa
a(t) &=& a_{0} + a_{2}\,t^{2}  + \text{O}(t^{4})\,,
\\[2mm]
a_{0} &>& 0\,,
\\[2mm]
a_{2} &>& 0\,,
\eeqa
\esubeqs
the connection components \eqref{eq:RWK-connection-m-0}
are well-behaved at $t=0$,  finite in fact.
As will be seen in
Sec.~\ref{subsec:Defect cosmological solution without curvature singularity},
the reduced Einstein equations give precisely
this bounce-type behavior of $a(t)$.
Incidentally, an odd function $a(t)=-a(-t)$ for $t \ne 0$ with
a discontinuity $a=\pm \, a_{0}$ at $t=0$ would give
an ill-defined (possibly infinite) term from  the $\dot{a}$ factor
in the spin connection components \eqref{eq:RWK-connection-m-0}.

We have a further remark on the structure of
the tetrad \eqref{eq:RWK-tetrad}.
Using instead a different tetrad (marked by a tilde),
\bsubeqs\label{eq:RWK-tetrad-BAD}
\beqa
\label{eq:RWK-tetrad-a-is-0-BAD}
\widetilde{e}^{0}
&=&
\sqrt{\frac{t^{2}}{t^{2}+b^{2}}}\;\text{d}t
=
\frac{|t|}{\sqrt{t^{2}+b^{2}}}\;\text{d}t\,,
\\[2mm]
\label{eq:RWK-tetrad-a-is-m-BAD}
\widetilde{e}^{m}
&=&  a(t) \,\text{d}x^{m}\,,
\eeqa
we obtain the following spin connection components:
\beqa
\label{eq:RWK-connection-m-0-BAD}
\widetilde{\omega}^{m}_{\phantom{z}0}
&=&
-\widetilde{\omega}^{0}_{\phantom{z}m}
=
\frac{\sqrt{t^{2}+b^{2}}}{|t|}\;        \dot{a}\; \text{d}x^{m}\,.
\eeqa
With bounce-type behavior \eqref{eq:bounce-type-a},
there is then a $t/|t|$ discontinuity at $t=0$ in these
spin connection components, which gives a delta-function
singularity in the corresponding curvature 2-form components,
\beqa
\label{eq:RWK-curvature-m-0-BAD}
\widetilde{R}^{m}_{\phantom{z}0}
&=&
-\widetilde{R}^{0}_{\phantom{z}m}
=
4\,a_{2}\,b\;\delta(t)\;dt \wedge \text{d}x^{m} + \ldots \,.
\eeqa
\esubeqs
A similar observation can be made for
the defect-wormhole solution, as will be discussed in
\ref{subapp:Defect-wormhole tetrads and spin connection}.

For the actual tetrad choice \eqref{eq:RWK-tetrad} and
an even function $a(t)$, the spin connection components
$\omega^{\phantom{z}a}_{\mu\phantom{z}b}(x)$
from \eqref{eq:RWK-connection} are even under $t \to -t$,
while the tetrad component $e^{0}_{\phantom{z}\mu}$
from \eqref{eq:RWK-tetrad-a-is-0} is odd
and the tetrad components
$e^{m}_{\phantom{z}\mu}$ from \eqref{eq:RWK-tetrad-a-is-m} are even.
This T-odd behavior of the tetrad $e^{0}_{\phantom{z}\mu}$
(without parity reversal of the spatial components
$e^{m}_{\phantom{z}\mu}$) differs from the CPT-odd behavior
$e^{a}_{\phantom{z}\mu}(t,\,x^{m})  = - e^{a}_{\phantom{z}\mu}(-t,\,x^{m})$
discussed in Ref.~\refcite{BoyleFinnTurok2018},
which has, however, a divergent curvature at $t=0$.

%%%%\newpage%%tmp
\subsection{Defect cosmological solution without curvature singularity}
\label{subsec:Defect cosmological solution without curvature singularity}

Having obtained modified Friedmann equations, it is clear that we
expect to get modified solutions. In fact,
for constant EOS parameter $w_{M} \equiv P_{M}/\rho_{M} =1/3$,
the solution of \eqref{eq:mod-Feqs}
reads~\cite{Klinkhamer2019-prd}
\bsubeqs\label{eq:regularized-Friedmann-asol-rhoMsol}
\beqa\label{eq:regularized-Friedmann-asol}
a(t)\,\Big|_{(w_{M}=1/3)}^\text{(FLRWK)}
&=&
\sqrt[4]{\big(t^{2}+b^{2}\big)\big/
\big(t_{0}^{2}+b^{2}\big)}\,,
%%\eeqa
\\[4mm]
%%\beqa
\label{eq:regularized-Friedmann-rhoMsol}
\rho_{M}(t)\,\Big|^\text{(FLRWK)}_{(w_{M}=1/3)} &=&
\rho_{M0}/a^{4}(t) =\rho_{M0}\,
\big(t_{0}^{2}+b^{2}\big)\big/\big(t^{2}+b^{2}\big) \,,
\eeqa
\esubeqs
with $a(t_{0})=1$ at $t_{0} \geq 0$ and  $\rho_{M0}>0$
[actually, $G \rho_{M0} \propto 1/(b^{2}+t_0^{2})\,$,
as follows from \eqref{eq:mod-1stFeq}].

\begin{figure}[t]%[h]
\begin{center}
\hspace*{0mm}   %%renamed FIG2-v1.eps=FIG02-v4.eps
\includegraphics[width=0.55\textwidth]{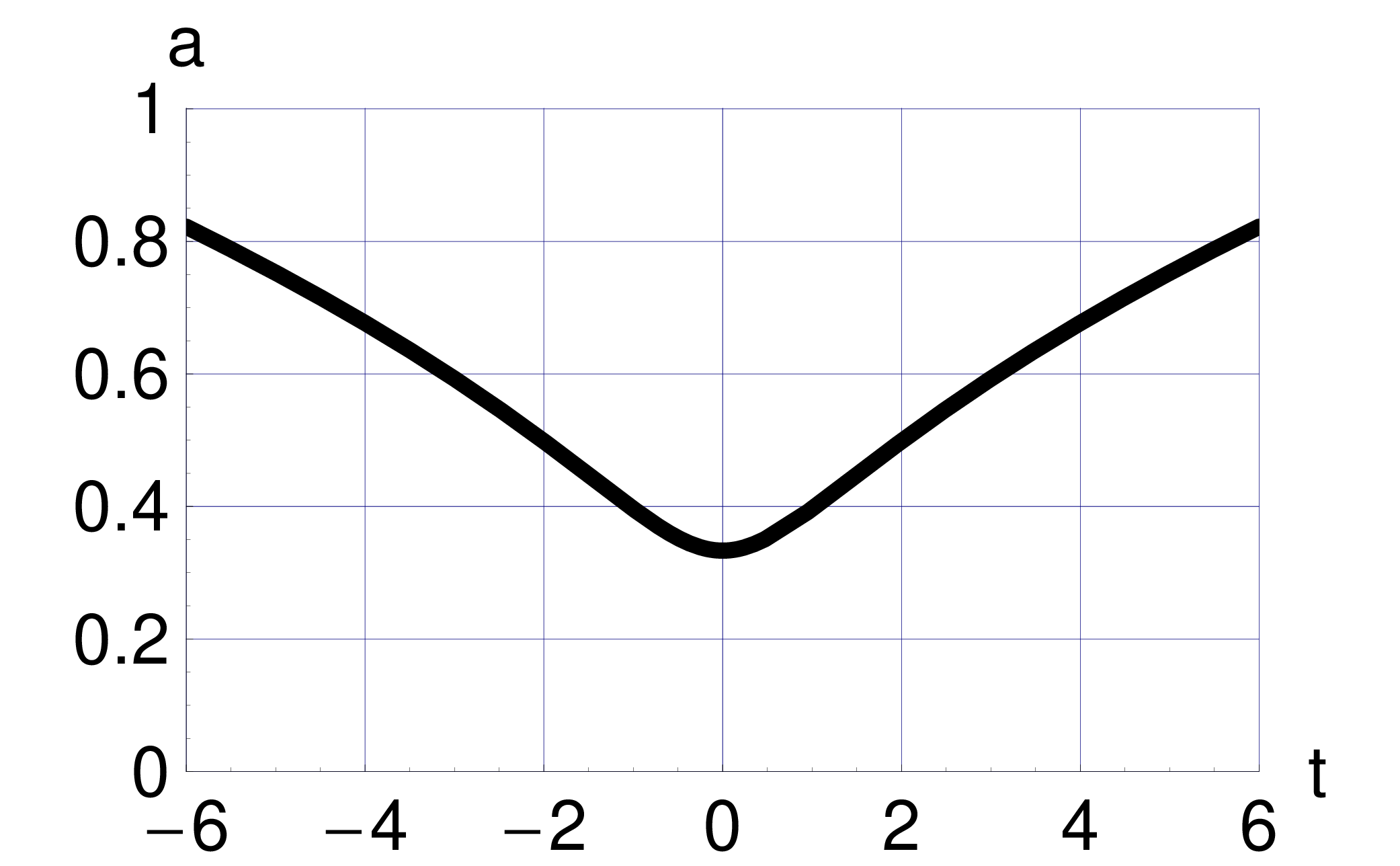}
%%{BB-as-spacetime-defect-FIG2-v012.eps}
%{Proc-Cracow-Epiphany-2021-Klinkhamer-fig1-v1.eps}
\end{center}
\vspace*{8pt}
\caption{Cosmic scale factor $a(t)$ of the spatially-flat
FLRWK universe with $w=1/3$ matter,
as given by \eqref{eq:regularized-Friedmann-asol}
for $b=1$ and $t_{0}=4\,\sqrt{5}$.
\vspace*{-0mm}}
\protect\label{fig:a-FLRWK}
\end{figure}

The new solution \eqref{eq:regularized-Friedmann-asol-rhoMsol}
is perfectly smooth at $t=0$ as long as $b\ne 0$
(cf. Fig.~\ref{fig:a-FLRWK}).
The same holds for the corresponding
Kretschmann curvature scalar
$K\equiv R^{\mu\nu\rho\sigma}\,R_{\mu\nu\rho\sigma}$,
\beq
\label{eq:FLRWK-sol-K-regular}
K\,\Big|^\text{(FLRWK)}_{(w_{M}=1/3)}
\propto 1\big/\big(b^{2}+t^{2}\big)^{2}\,,
\eeq
which is finite at $t=0$ as long as $b\ne 0$.

Observe that the function $a(t)$ from \eqref{eq:regularized-Friedmann-asol}
is \emph{convex} over a finite interval around $t=0$
(see Fig.~\ref{fig:a-FLRWK}), whereas
the function $a(\tau)$ from \eqref{eq:Friedmann-asol},
with $t$ replaced by $\tau$ and $t_{0}$ by $\tau_{0}$,
is \emph{concave} for $\tau \geq b >0$ (cf. Fig.~\ref{fig:a-FLRW}).
This different behavior of $a(t)$ just above $t=0$ (convex)
and of $a(\tau)$ just above $\tau=b$ (concave)
results from the different differential structures mentioned
%in Sec.~\ref{subsec:Defect cosmological metric and modified Friedmann equations}.
in the text under \eqref{eq:RW-ds2-for-tau-coord}
and the second remark under \eqref{eq:mod-Feq-EOS}.

%Returning to the pure-time defect of
%the cosmological metric \eqref{eq:RWK-ds2}, we refer to the
Various aspects of this defect cosmology have been discussed
in the follow-up papers~\cite{Klinkhamer2020-more,KlinkhamerWang2019-cosm,%
KlinkhamerWang2020-pert,Wang2021,Battista2021}
and the review~\cite{Klinkhamer-Epiphany2021}.
The papers~\cite{Wang2021,Battista2021} address, in particular,
the  discontinuous behavior at $t=0$, for example the  discontinuity
in the extrinsic curvature of constant $t$ hypersurfaces.

Degenerate-metric cosmologies have also been studied
in Ref.~\refcite{Holdom2023}.
The metric \emph{Ansatz} (1) of that paper
for $D=1$, $\ell=1$, $u=0$, and $d(t)=\sqrt{1+t^{2}}$
reproduces precisely our defect-cosmology metric \eqref{eq:RWK-ds2}
for $b=1$ and the case of relativistic matter ($w_{M}=1/3$) with
$a(t)=(1+t^{2})^{1/4}$.
Hence, part of the (formal) analysis of Ref.~\refcite{Holdom2023}
carries over to our case, in particular the final
expressions for the Einstein tensor
$G_{\mu\nu}$ in Eq.~(3) of that paper and the curvature
invariants $R$ and $K$ in Eq.~(5) of that paper,
without singularities whatsoever.

%%%%%%\newpage%%tmp
\subsection{Mathematics: Continuous extension}
\label{subsec:Mathematics-Continuous-extension}

We have thus seen that 
the nonsingular equations (\ref{eq:Feqs}abcd) have  
a singular solution \eqref{eq:Friedmann-solution},
while 
the singular equations (\eqref{eq:mod-Feqs}abcd) have  
a regular solution \eqref{eq:regularized-Friedmann-asol-rhoMsol}.

As mentioned below \eqref{eq:mod-Feqs}, these singular equations
are solved by a ``continuous-extension'' procedure.
In order to explain this point in more  detail,   
and as this review is pedagogical,
it suffices to present two quotes from Horowitz' 1991 paper.

The first quote is from Sec.~4, p. 598 in Ref.~\refcite{Horowitz1991}
(again with $g \equiv \det[g_{\mu\nu}]$, but
changing the equation number there to the one relevant here):\newline  
``($\ldots$) there are many examples of smooth solutions to singular equations.
Bessel’s equation, for example, has a regular singular point at the origin,
but the Bessel functions are smooth solutions everywhere. Similarly,
a smooth $g_{\mu\nu}$, for which $g = 0$ on a set of measure zero,
and $G_{\mu\nu}= 0$ whenever $g \ne 0$ could be considered a solution to the vacuum Einstein equation everywhere. This is because Einstein’s equation takes
the form
\beq
\label{eq:Horowitz1991-Eq.4.3}
W_{\mu\nu}(g_{\alpha\beta})/g^2 = 0 \,, %%\qquad (4.3)   \nonumber
\eeq
where $W_{\mu\nu}$ is a continuous function of
$g_{\alpha\beta}$ and its first two derivatives (but not its
inverse). If this holds for a metric $g_{\mu\nu}$,
with $g \ne 0$ almost everywhere, then $W_{\mu\nu} =0$
almost everywhere. If $g_{\mu\nu}$ is smooth, then
$W_{\mu\nu} = 0$ everywhere. (There is no
possibility of a $\delta$-function type contribution arising from a
$C^\infty$ $g_{\mu\nu}$ and its derivatives.)''

The second quote is from Sec.~4, p. 599 in Ref.~\refcite{Horowitz1991}
(here, $\mathcal{S}$ is a compact 3-manifold and $M$ the four-dimensional
spacetime manifold considered):\newline
``It is now straightforward to extend this discussion to include matter fields (which
are described by covariant tensors). One can start with a smooth solution to the
coupled Einstein-matter field equations on
$\mathcal{S} \times \mathbb{R}$ and pull back the metric $g_{\mu\nu}$ and
matter fields under a smooth map from $M$ to $\mathcal{S} \times \mathbb{R}$. 
At all points where $g \ne 0$, the
fields on $M$  satisfy the coupled Einstein-matter field equations.
Furthermore, $g \ne 0$
almost everywhere and all fields are smooth. Hence they are solutions everywhere.
Note that all scalars remain finite as the metric becomes degenerate.
For example, in
the case of a scalar field the appropriate components of $\nabla_\mu \phi$
vanish at precisely the
points where the metric becomes degenerate so that
$g^{\mu\nu} \nabla_\mu \phi \nabla_\nu \phi$  remains finite.''

%%%%\newpage%%tmp
A few remarks are in order:
\begin{enumerate}
  \item
The $R_{\mu\nu} = W_{\mu\nu}/g^2$ observation of the first quote
has already been made by Einstein and Rosen in the paragraph starting
with ``We now ask $\ldots$''  at the bottom of the left column
on p.~74 of Ref.~\refcite{EinsteinRosen1935}.
One possible derivation of this identity,
referring to equation numbers in Ref.~\refcite{Weinberg1972},
starts from the
expression (6.5.2) for the commutator of covariant derivatives
on an arbitrary contravariant \mbox{4-vector},
contracts the indices $\lambda$ and $\nu$,
and uses the expression (4.7.7) for the covariant divergence.
\item
The mathematics of the continuous-extension procedure
has also been detailed
in Sec.~3.3.1 of Guenther's Master Thesis~\cite{Guenther2017}.
  \item
A crucial point in the first quote
is to have an infinitely differentiable metric $g_{\mu\nu}$,
so that there arise no delta-function type contributions in, for example, the
curvature scalars.
\end{enumerate}

Expanding on the last remark,
examples of pure-space defects with nondifferentiable metric
components and delta-function contributions to the curvature
have been given by Schwartz in his
PhD Thesis~\cite{Schwarz2010}. The simplest example
(on p.~35 of that reference) has a 2-dimensional metric given by
\bsubeqs\label{eq:2-dimensional-example}
\beqa
\label{eq:2-dimensional-example-ds2}
ds^{2}  &=&  dY^{2} + \Big( b + |Y| \Big)^{2}\, dX^{2}\,,
\eeqa
with spatial coordinates $X$ and $Y$ ranging
over $(-\infty,\,\infty)$ and a resulting Ricci curvature
\beqa
\label{eq:2-dimensional-example-R}
R  &=&  - \frac{4}{b}\;\delta(Y)\,,
\eeqa
\esubeqs
which is infinite at the defect line $Y=0$.
The origin of the problem lies in the nonanalytic behavior
of the metric component $g_{XX}$ from \eqref{eq:2-dimensional-example-ds2},
which has a nondifferentiable contribution $2\, b\, |Y|$.
In our case, we postulate a $C^\infty$ metric $g_{\mu\nu}$, as mentioned in
Horowitz' first quote and our third remark.

%%%%\newpage%%tmp
\subsection{Singularity theorems}
\label{subsec:Singularity theorems}

There is no doubt as to the validity of
the Hawking and Hawking--Penrose cosmological
singularity theorems~\cite{Hawking1967,HawkingPenrose1970}, but
the ``singularity'' of these theorems need not correspond to
a curvature singularity and
may very well correspond to a lower-dimensional spacetime defect
with a locally degenerate metric.  Hawking states this explicitly
on pp.~188--189 in Sec.~1 of Ref.~\refcite{Hawking1967}:\newline
``This brings us to the third question: the nature of the singularity.
In fact, what the various theorems actually prove is that
the space-time manifold cannot be timelike and null geodesically complete with a $C^2$ metric. The reason for adopting this as the definition of a singularity is as follows. If there was a point of the space-time manifold at which the metric was degenerate or not $C^2$, we could say that this was a singularity. However, $\ldots$''
The rest of the quote will be given at the end of this subsection.

The same statement has also been made by Horowitz and
the relevant quote is from Sec.~4, p. 599 in Ref.~\refcite{Horowitz1991}
(changing the reference number there to the one2 used here):\newline
``However the singularity theorems~\cite{HawkingPenrose1970} %%[27]
show that most solutions on $\mathcal{S} \times \mathbb{R}$
are geodesically incomplete. This is
usually interpreted as evidence for unbounded curvature resulting
from gravitational collapse. However, in some cases the geodesic
incompleteness is a sign that the metric becomes degenerate
but the curvature remains finite.''

%
%which contains an interesting
%discussion on the question of how to deal with these singular points.
%The rest of the quote will be given at the end of this subsection,
%as it deals with another issue.
%

For later reference,
we give now already the rest of Hawking's quote
%addressing the question of how to deal with these singular points
(from p.~189 in Sec.~1 of Ref.~\refcite{Hawking1967}):
``However, we could cut out the singular points and say that
the remaining manifold represented all of space-time.
Indeed, it would seem undesirable to include the singular points
in the definition of space-time, as, if we
did, we would be introducing
something into the theory which was not
physically observable; namely, the
manifold structure (i.e. the admissible
coordinates) and the metric at those
points. On the other hand, we want to
omit only the singular points and not
perfectly regular points as well.
%%part quote omitted:
%``We shall therefore adopt the Postulate of Inextendability.
%By this we mean that the space-time manifold,
%$\mathcal{M}$, is not a proper open submanifold
%of a four-dimensional, connected, paracompact, $C^{3}$ manifold
%on which there is a $C^{2}$
%Lorentz metric which coincides with the
%physical metric on $\mathcal{M}$.''
($\dots$)  Although we have omitted the singular points
from the definition of space-time, we may still be able to recognize the `holes' left
where they have been cut out by the existence of geodesics
which cannot be extended to arbitrary values of the affine parameter.''
Most likely,  the proper treatment of these
troublesome singular points will only be resolved
when the \emph{origin} of these special points has been
established (see also Chap.~8 of the
monograph~\cite{HawkingEllis1973} for further discussion).

In this section, we have introduced
what we may call ``defect cosmology.''
The next two sections address two follow-up questions.
The first section (Sec.~\ref{sec:Communication-between-the-two-worlds})
is of a more phenomenological  nature
(addressing the issue of communication between the two branches
of the full cosmological solution)
and the second section (Sec.~\ref{sec:Nature-and-origin})
is more fundamental (considering the possible
origin of this particular spacetime defect).

%%\newpage%%tmp
\section{Communication between the two worlds}
\label{sec:Communication-between-the-two-worlds}

%\newpage
\subsection{Coordinate time and thermodynamic time}
\label{subsec:Coordinate time and thermodynamic time}

The spacetime metric \eqref{eq:RWK-ds2}
with coordinates \eqref{eq:RWK-ranges-coordinates} for
$b^{2} = 0$ and $a(t) =1$ formally
reproduces the standard Minkowski metric which solves
the Einstein equation for the case of vanishing matter
content and zero cosmological constant $\Lambda$.
The causal structure of the defect spacetime
\eqref{eq:RWK} with $b^{2} \ne 0$ is then,
by continuity, the same as that of the
standard Minkowski spacetime and the corresponding
Penrose conformal diagram will be given later
in Sec.~\ref{subsec:Direct communication or not}.

For the spacetime metric \eqref{eq:RWK} and a homogeneous
ultrarelativistic perfect fluid ($w_{M}=1/3$),
we have obtained the $a(t)$ and $\rho_{M}(t)$ solutions
\eqref{eq:regularized-Friedmann-asol}
and \eqref{eq:regularized-Friedmann-rhoMsol}.
In order to simplify the discussion later on, we now take
a \emph{single} homogeneous pressureless perfect fluid ($w_{M}=0$)
instead of a mix of two or more fluids,
\bsubeqs\label{eq:regularized-Friedmann-asol-rhoMsol-wMzero}
\beqa\label{eq:regularized-Friedmann-asol-wMzero}
a(t)\,\Big|_{(w_{M}=0)}^\text{(FLRWK)}
&=&
\sqrt[3]{\big(b^{2}+ t^{2}\big)\big/b^{2}}\,,
\eeqa
%%\\[4mm]
\beqa
\label{eq:regularized-Friedmann-rhoMsol-wMzero}
\overline{\rho}_{M}(t)\,\Big|^\text{(FLRWK)}_{(w_{M}=0)} &=&
\overline{\rho}_\text{bounce}/a^{3}(t) =
\overline{\rho}_\text{bounce}\,b^{2}\big/\big(b^{2}+ t^{2}\big) \,,
\eeqa
\esubeqs
where $a(t)$ is normalized to unity at $t=0$
and the bar on $\rho$ emphasizes that this is
the homogeneous (unperturbed) component.

Indeed, scalar metric perturbations for this spacetime
 have been studied in Ref.~\refcite{KlinkhamerWang2020-pert}.
It was found that a plane-wave scalar metric
perturbation (wave vector $\mathbf{k}$),
with a short wavelength compared to the Hubble radius
$1/H \equiv a/\dot{a}$, behaves as follows,
in an abbreviated notation displaying only the time dependence:
\beq\label{eq:short-wavelength-sol.density-pert}
\left.\frac{\delta \rho_\mathbf{k}( t )}{\overline{\rho}( t )} 
\right|^\text{(short-wavelength)}
\sim
\widetilde{C}_{\mathbf{k},\,1}\, \Big( b^{2}+ t^{2} \Big)^{1/3} +
\widetilde{C}_{\mathbf{k},\,2}\, \Big( b^{2}+ t^{2} \Big)^{-1/2}\,,
\eeq
with arbitrary constants $\widetilde{C}_{\mathbf{k},\,1}$
and $\widetilde{C}_{\mathbf{k},\,2}$
depending on the boundary conditions;
the full result is given by
Eq.~(3.22b) in Ref.~\refcite{KlinkhamerWang2020-pert},
with the corresponding result for
the scalar metric perturbations in Eq.~(3.19) there.
Observe that the growing part
in \eqref{eq:short-wavelength-sol.density-pert}
is proportional to $a(t)$.

The mathematics of the
solution \eqref{eq:regularized-Friedmann-asol-rhoMsol-wMzero}
and the perturbations \eqref{eq:short-wavelength-sol.density-pert}
is perfectly clear,
but its physical interpretation is less so.
In fact, there appear to be two different physical
interpretations.

First, we interpret the spacetime from \eqref{eq:RWK}
and \eqref{eq:regularized-Friedmann-asol-rhoMsol-wMzero}
as a standard bounce
solution~\cite{IjjasSteinhardt2018,Brandenberger2023},
having a contracting universe $U_{-}$ for $t<0$
(pre-Big-Bang phase) and
an expanding universe $U_{+}$ for $t>0$
(post-Big-Bang phase),
with a bounce at $t=0$ where $da/dt = 0$.
Here, the coordinate time $t$ is interpreted as the
``physical'' time.

%%%%\newpage%%tmp
Second, we interpret the physical time as
the ``thermodynamic'' time $\mathcal{T}$
for which matter density perturbations grow
(see the discussion on p. 2, left column
of Ref.~\refcite{BoyleFinnTurok2018}
and in the last paragraph of Sec.~IV of
Ref.~\refcite{KlinkhamerWang2020-pert}).
With the result \eqref{eq:short-wavelength-sol.density-pert},
we have the following relation
between this thermodynamic time $\mathcal{T} \in [0,\,\infty)$
and the coordinate time $t \in (-\infty,\,\infty)$:
\beq
\label{eq:thermodynamic-time-coordinate}
\mathcal{T}
\equiv
\begin{cases}
 +t \,,   &   \text{for}\quad t \geq 0 \,,
 \\[2mm]
 -t \,,   &  \text{for}\quad t\leq 0\,.
\end{cases}
\eeq
We can then rewrite \eqref{eq:regularized-Friedmann-asol-wMzero} as
\beq
\label{eq:regularized-Friedmann-asol-wMzero-mathcalT}
a(\mathcal{T})\,\Big|_{(w_{M}=0)}^\text{(FLRWK)}
=
\sqrt[3]{\big(b^{2}+ \mathcal{T}^{2}\big)\big/b^{2}}\,,
\eeq
so that we have \emph{two} expanding worlds $W_{\pm}$
for $\mathcal{T} >0$,
corresponding to world $W_{+}$ for $t>0$
and to world $W_{-}$  for $t<0$.
Incidentally, we prefer the use of the word ``world''
and its plural for etymological reasons (strictly speaking,
the word ``universe'' refers to a single entity, the
whole of space and time and matter).

For this second interpretation and
following the suggestion of Ref.~\refcite{BoyleFinnTurok2018},
we may consider both worlds $W_{+}$ and $W_{-}$
to be ``pair created'' at $\mathcal{T}=0$ and
these worlds may be called the two
different ``sides'' of the Big Bang.
In our case as it stands,
the other side ($W_{-}$) of the Big Bang is not
the parity reversal of our
side $W_{+}$ (and $W_{-}$ may also not have antiparticles
corresponding to the particles of $W_{+}$), so that
the term ``pair creation'' should not be taken literally
and we prefer to speak of the ``emergence'' of the
worlds $W_{+}$ and $W_{-}$ at $\mathcal{T}=0$.

%quote from {BoyleFinnTurok2018}:
%``so that our CPT-invariant
%Universe is reinterpreted as a universe-antiuniverse pair
%($U\bar{U}$), emerging from nothing.''

%%%%\newpage%%tmp
\subsection{Direct communication or not}
\label{subsec:Direct communication or not}

We now turn to the issue of possible physical
communication between the two worlds $W_{\pm}$
(specifically, sending messages between them).
This depends on the physical interpretation of
the mathematical solution, as discussed in
Sec.~\ref{subsec:Coordinate time and thermodynamic time}.

In the first interpretation of a standard bounce cosmology
(with a contracting pre-Big-Bang phase for $t<0$
and an expanding post-Big-Bang phase for $t>0$),
we can, in principle, have a gravitational wave  traveling
from the pre-Big-Bang phase to the post-Big-Bang phase
(cf. App.~B~2 of Ref.~\refcite{KlinkhamerWang2020-pert}).
The wave would have emission time
$t_\text{em}<0$  and observation time $t_\text{obs}>0$
(see the next paragraph for further discussion using the
relevant Penrose conformal diagram).
If these gravitational waves were emitted by known sources
(e.g., collapsing binary black holes), then they
could be considered as ``standard sirens''
(analogous to ``standard candles'' for electromagnetic radiation);
cf. Sec.~19.6 in Ref.~\refcite{Maggiore2018} for further discussion.
Such standard sirens
would, however, only produce barely audible ``songs''  to us
(very weak signals, in less poetic language),  because these
sources would have
%light-travel times of more than $10^{10}$ years.
distances of more than $10^{10}$ lightyears.
Assuming it to be possible to determine a corresponding
redshift $z$ given by $a(t_\text{obs})/a(t_\text{em})-1$,
these weak  signals could, in principle, contribute a second curve
in the Hubble diagram with unusual redshifts/blueshifts,
in addition to the usual curve from standard sirens
in our post-Big-Bang world, as discussed in
Secs.~III~C and D of Ref.~\refcite{KlinkhamerWang2019-cosm}.

\begin{figure}[t]   %%[p]   %%
\vspace*{0mm} %%renamed FIG03-v4.eps
\centerline{
\includegraphics[width=0.3\textwidth]{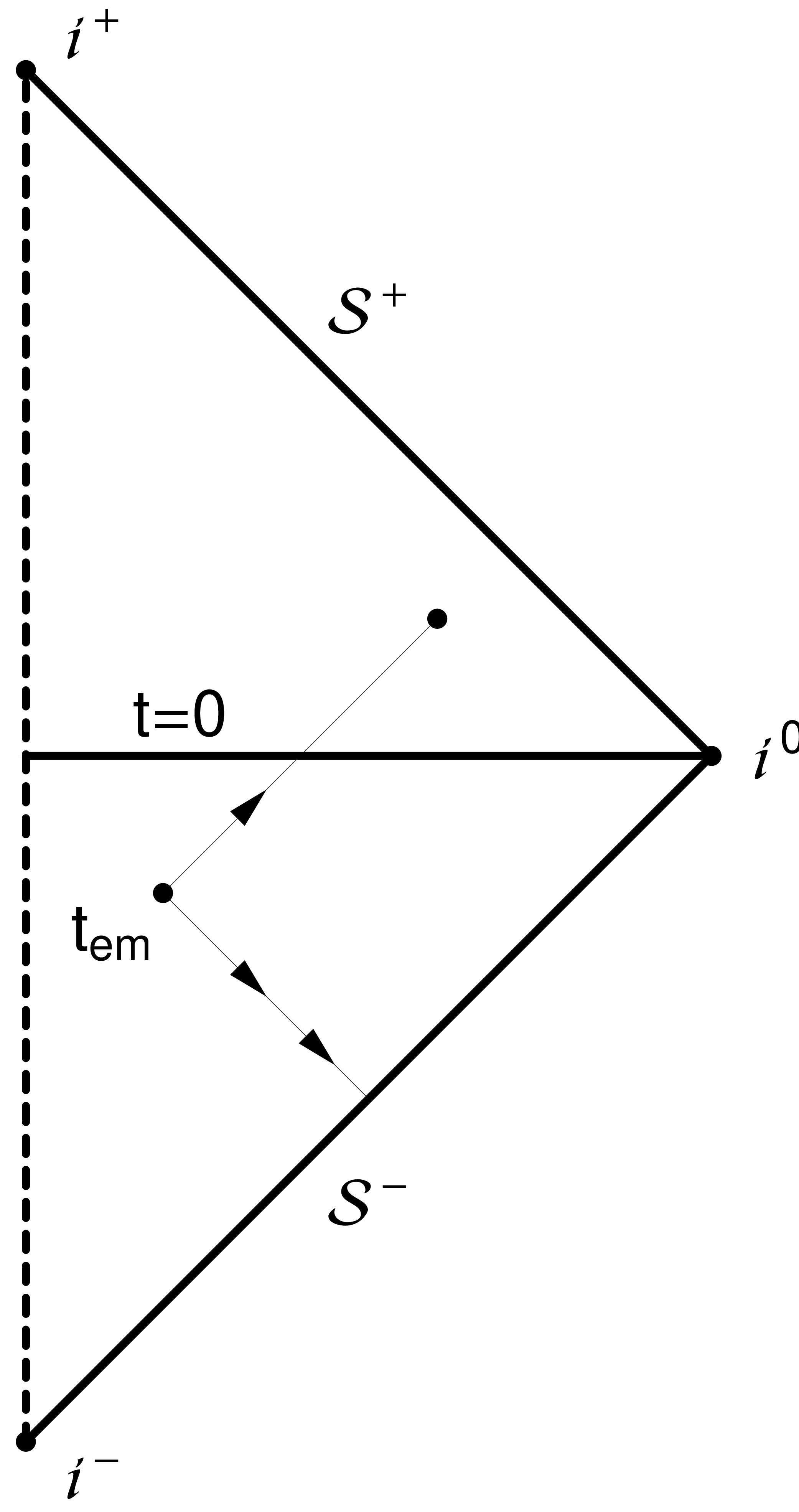}}
%%{BB-as-spacetime-defect-FIG4-v270.eps}
\vspace*{8pt}
\caption{Penrose conformal diagram of the defect-cosmology
spacetime \eqref{eq:RWK} with scaling function $a(t)$
from \eqref{eq:regularized-Friedmann-asol-wMzero},
using the same notation as in Fig.~15(ii) of
Ref.~\protect\refcite{HawkingEllis1973}
and Fig.~27.16b of Ref.~\protect\refcite{Penrose2005}.
One gravitational wave with emission time
$t_\text{em}<0$  and tentative observation time $t_\text{obs}>0$
is shown as the thin curve with a single arrow going diagonally up
and another wave emitted at $t_\text{em}<0$
as the thin curve with two arrows going diagonally down;
see the text for the explanation that, under certain assumptions,
only the last gravitational wave will be emitted physically.}
\protect\label{fig:Penrose-diagram-defect-WH}
\vspace*{0mm}
\end{figure}

%%\newpage%%tmp
The Penrose conformal diagram of the defect-cosmology spacetime
(Fig.~\ref{fig:Penrose-diagram-defect-WH})
has, as mentioned before,
the same structure as that of Minkowski spacetime
(a diamond-shaped diagram shown as Fig.~15(ii) in
Ref.~\refcite{HawkingEllis1973}
and Fig.~27.16b in Ref.~\refcite{Penrose2005}).
According to the surgery construction of Sec.~3 in
Ref.~\refcite{Klinkhamer2019-prd}, the diagram
of Fig.~\ref{fig:Penrose-diagram-defect-WH} can be obtained
from an expanding-RW-universe  triangular diagram truncated
to $t \geq b$ (the full diagram is shown in Fig. 21(iii)
in Ref.~\refcite{HawkingEllis1973}
and Fig.~27.17b in Ref.~\refcite{Penrose2005})
and a contracting-RW-universe  triangular diagram truncated
to $t \leq -b$,
which are glued together at $t=\pm b$.
Figure~\ref{fig:Penrose-diagram-defect-WH}  also
shows, as the thin curve with a single arrow,
a gravitational (retarded) wave  emitted in the pre-Big-Bang phase
and observed in the post-Big-Bang phase.
Everything appears to be fine with causality,
as long as $t$ is the physical time in the emission process.

%%https://image.online-convert.com/convert/jpeg-to-eps  CUTOFF   tmp1
%%https://convertio.co/jpeg-eps/  GRID   tmp2

But, in the second interpretation with physical
(thermodynamic) time $\mathcal{T} = |t|$, we will argue
that the retarded classical wave solution as discussed
above is physically unrealistic in the  $t<0$ phase.
Mathematically, the gravitational wave shown
in Fig.~\ref{fig:Penrose-diagram-defect-WH}
as the thin curve with a single arrow corresponds to
a retarded solution from a source at $t_\text{em}<0$.
More precisely, it is retarded in terms of the coordinate time $t$
and, for clarity, we will speak of $t$-retarded.
Now, the point is that this gravitational wave
would be \mbox{$t$-retarded} but $\mathcal{T}$-\emph{advanced} initially.
In a more or less standard matter-dominated world $W_{-}$
with $\mathcal{T}$-expansion and
matter density perturbations increasing with positive $\mathcal{T}$
(see further discussion below),
this $\mathcal{T}$-advanced gravitational wave in the
second world $W_{-}$ would be completely unrealistic  %%highly unusual
(the electromagnetic-wave discussion in Chapter 5 of
Ref.~\refcite{Davies1977} carries over to gravitational waves).
A possible $\mathcal{T}$-\emph{retarded} wave emitted at
$\mathcal{T}_\text{em}=|t_\text{em}|$ for
$t_\text{em}<0$ would be more realistic and
is shown, in Fig.~\ref{fig:Penrose-diagram-defect-WH},
as the thin curve with two arrows running down diagonally towards
the infinity surface $\mathcal{S}^{-}$
(with $r=\infty$, $\mathcal{T}=\infty$, and $t=-\infty$).
Hence, gravitational waves emitted by 
localized sources in such a world $W_{-}$   
cannot be expected, under realistic emission conditions, to be such as to reach
the world $W_{+}$.

%%%%%%%%%%%%%%%%%%\newpage%%tmp
Let us present a brief intermezzo with some details
on the argument in favor of
$\mathcal{T}$-retarded electromagnetic or gravitational waves.
The issue is really the choice of boundary conditions appropriate
to the physical situation of the localized emission sources.
Start with the $W_{+}$ world (ours by definition), which is known
to expand with $\mathcal{T}=t$ for positive $t$
and to be near-homogeneous initially
($\mathcal{T}=t$ being perhaps of the order of the Planck time),
with perturbations growing as time $\mathcal{T}=t$ advances
(for now, we will drop the mention of $t$ and stick with $\mathcal{T}$).
At a certain moment, stars and black holes are formed with
the possibility of providing standard sources
at $\mathcal{T}_\text{em} >0$
(for gravitational waves, the prime example being
merging binary black holes).
According to all of our experience,
the electromagnetic/gravitational radiation emitted
is \emph{known} to be of the
\emph{retarded} variety with respect to time $\mathcal{T}$.
Simply put, the fields have zero field values and
first derivatives (the wave equation being second order)
at early times $\mathcal{T} < \mathcal{T}_\text{em}$
(cf. Sec.~5.1 of Ref.~\refcite{Davies1977}
and Sec.~6.4 of Ref.~\refcite{Jackson1999}).
%% Sec.~6.6 of Ref.~\refcite{Jackson1975}).
%
%quote
%The choice of A and B in (6.61) depends on the
%boundary conditions in time that specify the
%physical problem. It is intuitively obvious that,
%if a source is quiescent until some time t=0 and
%then begins to function, the appropriate Green
%function is the first term in (6.39).  %%(6.61),
%corresponding to waves radiated outwards from the
%source after it begins to work.
%

Indeed,   %%EE
the choice between retarded and advanced Green's functions
``depends on the \emph{boundary conditions in time}
that specify the physical problem.
It is intuitively obvious that, if a source is quiescent 
until some time $\mathcal{T} =\mathcal{T} _\text{em}$ and then begins to function,
the appropriate Green function is
the retarded Green's function,
corresponding to waves radiated outwards from the source after it begins to work''
(the quote is from page~244 in Sec.~6.4 of Ref.~\refcite{Jackson1999},
where, in the second sentence,
we have replaced ``$t=0$'' by ``$\mathcal{T} =\mathcal{T} _\text{em}$''
and ``the first term in (6.39)'' %%(6.61)''
by ``the retarded Green's function,''
in order to get a sentence relevant to our context).   
Hence, for times after emission, we have only $\mathcal{T}$-retarded waves
propagating in our ($W_{+}$) world.

Turning to the $W_{-}$ world (not ours but another world!),
there is also expansion \eqref{eq:regularized-Friedmann-asol-wMzero-mathcalT} 
with $\mathcal{T}=-t>0$ for negative $t$
(from now on, we drop the mention of the negative $t$
and stick with the positive $\mathcal{T}$).
In line with the ``emergence'' scenario sketched in the last paragraph
of Sec.~\ref{subsec:Coordinate time and thermodynamic time},
we assume that the $W_{-}$ world is also
near-homogeneous initially and has
perturbations growing as time $\mathcal{T}$ advances,
according to \eqref{eq:short-wavelength-sol.density-pert}
with $t^{2}$ replaced by $\mathcal{T}^{2}$.
Perhaps, at a certain moment, stars and black holes
are formed with the possibility of giving
sources at $\mathcal{T}_\text{em} >0$.
%(for gravitational waves, the prime example being
%merging binary black holes).
The electromagnetic/gravitational radiation emitted
can then be expected to be of the
\emph{retarded} variety with respect to the relevant
time $\mathcal{T}$.
Simply put, the fields have zero field values and
first derivatives (the wave equation being second order)
at early times $\mathcal{T} < \mathcal{T}_\text{em}$.
Hence, for times after
emission, we expect to have only $\mathcal{T}$-retarded
waves in the $W_{-}$ world, for example
the thin curve with two arrows
in Fig.~\ref{fig:Penrose-diagram-defect-WH}
and not the thin curve with a single arrow
(being $\mathcal{T}$-advanced).
This ends the intermezzo on
$\mathcal{T}$-retarded electromagnetic/gravitational waves.

%%%%%%%%%%%%%%%%\newpage%%tmp
Without direct classical communication between the two worlds
(granting certain assumptions about   
the other world $W_{-}$
such as growing matter density perturbations),
we can only think of quantum effects such as entanglement
to establish a connection between both worlds.
With the unknown ``emergence'' process of the
two worlds $W_{\pm}$ at $\mathcal{T}=0$
(cf. the last paragraph of 
Sec.~\ref{subsec:Coordinate time and thermodynamic time}
and Sec.~\ref{sec:Nature-and-origin} below), 
we can even imagine an ultimate EPR-type pair-creation
situation~\cite{EPR1935} with
a first photon emerging in $W_{+}$ and
a second  photon emerging in $W_{-}$, where the two photons
correspond to an entangled quantum state
(cf. Chapter 8 of Ref.~\refcite{Ghirardi2005}).

It appears, however, that quantum-entanglement correlations
cannot be used to send a message from one world to the
other: the arguments are essentially the same as those
against faster-than-light
communication from quantum entanglement~\cite{EberhardRoss1989}
(a clear discussion is given in Sec.~11.2  of
Ref.~\refcite{Ghirardi2005}).
Still, adapting the last sentence on p.~147
of Ref.~\refcite{EberhardRoss1989},
possible quantum-entanglement-type messages
between the two worlds $W_{\pm}$
``should not be looked for in theories that abide
with orthodox quantum field theory but in theories
that allow some deviations from it.''
%full quote from p.~147 of Ref.~\refcite{EberhardRoss1989}
%``Justification for any effect providing faster-than-light
%communication
%should not be looked for in theories that abide
%with orthodox quantum field theory but in theories that allow some
%deviations from it.''
%
And precisely such a novel theory may also be needed to explain  
the emergence of these two worlds $W_{\pm}$, as will be
discussed further in the second half of
Sec.~\ref{sec:Nature-and-origin}.

%

%%\newpage%%tmp
\section{Nature and origin of the Big Bang defect}
\label{sec:Nature-and-origin}

It perhaps needs to be emphasized that we are \emph{not} considering
\emph{standard} general relativity.
In fact, the degenerate metrics considered invalidate the standard
elementary flatness property precisely at the spacetime points of
the defect (a submanifold with vanishing determinant of the metric);
see App.~D of Ref.~\refcite{Klinkhamer2014-mpla}.
This implies that the spacetime points of the defect submanifold must be treated
differently than all other spacetime points (with nonvanishing metric
determinant) where the standard elementary flatness property does hold.
One possibility is to exclude
these troublesome points altogether from the
spacetime manifold, as suggested by Hawking in the quote of the
penultimate paragraph of Sec.~\ref{subsec:Singularity theorems}.
Another way to deal with these points
is by a continuous-extension procedure,
as explained by Horowitz in his 1991 paper~\cite{Horowitz1991}.
In this last approach, also adopted by us
(implicitly or explicitly), the
troublesome points are still considered to be part of spacetime
but are indeed treated differently, namely by continuous extension
(as reviewed
in our Sec.~\ref{subsec:Mathematics-Continuous-extension}).
Here, we are motivated by the known physical example of
atomic defects.

%At this point,
It may then be helpful to briefly recall the essence of
crystallographic defects in an atomic crystal.
The crystal defect is really an \emph{abstract} notion, namely a
discontinuous behavior of the crystal-ordering \emph{pattern} of
the atoms, and there is nothing ``wrong'' with the atoms themselves.
The relevant abstract mathematical spaces are the order-parameter
spaces, whose characteristics can be revealed by
a topological analysis; see Ref.~\refcite{Mermin1979}
for a comprehensive review with useful figures.
As to the \emph{origin} of the crystallographic defects,
we note that these defects are typically formed during
a \emph{rapid} crystallization process.

Returning to the Big-Bang spacetime defect as given in
Sec.~\ref{subsec:Defect cosmological solution without curvature singularity}, 
we observe that the Kretschmann curvature scalar $K$
and the matter energy density $\rho_{M}$
have the following orders of magnitude at $t=0$:
\bsubeqs\label{eq:K-rhoM-bounce}
\beqa
K\,\Big|^\text{(RWK)}_\text{defect}
&\sim& 1/b^4  \,,
\\[2mm]
\rho_{M}\,\Big|^\text{(RWK)}_\text{defect}
&\sim& E_\text{Planck}^{2} /b^{2}\,,
\eeqa
\esubeqs
with $b$ the length scale entering the metric \eqref{eq:RWK}
and $E_\text{Planck}$ the energy scale defined by $\sqrt{\hbar\,c^5/G}
\approx 1.22 \times 10^{19}\,\text{GeV}$.
As $E_\text{Planck}$ is the only realistic energy scale available,
the defect length scale $b$ will, without further input,
be proportional to the inverse of $E_\text{Planck}$,
so that both quantities in \eqref{eq:K-rhoM-bounce} are
of the order of $E_\text{Planck}^{4}$.

This last remark suggests that the origin of the
Big-Bang spacetime defect may be with a new quantum-gravity phase.
Heuristically, this new phase could contain ``atoms of space''
or even ``atoms of spacetime,''
which would crystallize (coagulate) into the classical spacetime as
described by general relativity.
The idea now is that this particular crystallization process
may very well be imperfect and that occasionally spacetime defects
remain.
If the emerging spacetime has a Lorentzian signature, then one
possible type of remnant defect could be the
Big-Bang spacetime defect discussed here.

%%%%%%%%%%%%%%%%%%%\newpage%%tmp
In order to do more than hand-waving, we need to understand
the mathematics and  physics  of such a new phase.
Nothing has been established definitely,
but one interesting suggestion relies on
superstring theory (cf. Ref.~\refcite{GreenSchwarzWitten1987}
with further references therein)
in the nonperturbative realization of a matrix model, of which
the IIB matrix model~\cite{IKKT-1997,Aoki-etal-review-1999}
is perhaps the most explicit example.

For this reason, we have reconsidered the question of how
precisely classical
spacetime would emerge from the IIB matrix model.
Loosely speaking, we have found that
the classical spacetime could result from
the basic structure (information content)
of the \emph{master-field} matrices~\cite{Klinkhamer2021-PTEP-master}.
The spacetime points would then emerge as certain averages of
their eigenvalues
and the inverse metric as the consequence of certain correlations in
the distributions of the extracted spacetime points.
A degenerate spacetime
metric could, in principle, arise~\cite{Klinkhamer2021-PTEP-big-bang}
from \emph{long-range tails}
of certain correlation functions contributing to
the emergent \emph{inverse} metric
(see also App.~C in Ref.~\refcite{Klinkhamer-Epiphany2021}).
Moreover, it is possible, in principle, to get
a Lorentzian signature from the master field of the well-defined
Euclidean IIB matrix model (the basic idea appears already in
Ref.~\refcite{Klinkhamer2021-PTEP-master} and has been
clarified by App.~D in Ref.~\refcite{Klinkhamer-Epiphany2021}).

The above discussion of a possible matrix-model origin
of the Big-Bang spacetime defect is, by necessity, concise.
An introductory review appears in Sec.~4 of
Ref.~\refcite{Klinkhamer-Epiphany2021} and
a more technical review in Ref.~\refcite{Klinkhamer-CORFU2021}.
In addition to the emergence of spacetime with an appropriate
defect, the novel theory also needs to explain the
emergence of matter, governed by standard quantum mechanics
or perhaps by quantum mechanics with (minor) modifications.

%%\newpage%%tmp
\section{Conclusion}
\label{sec:Conclusion}

%%BB of FLRW sol has metric eigenvalues of the strucure (-1000)
%%BB of defect sol has metric eigenvalues of the strucure (0111)

Degenerate metrics (interpreted as spacetimes
with localized defects)
have been used to ``tame'' potential
curvature singularities in three cases:
the curvature singularity 
at the center of a black hole~\cite{Klinkhamer2014-mpla},  
the curvature singularity
at the birth of the Universe~\cite{Klinkhamer2019-prd},  
and, more recently, 
the potential curvature singularity of a traversable
wormhole (references will be given in
\ref{app:Defect-wormhole solution}).
The actual physical properties at the center of a black hole
are shielded from us by the event horizon and the existence
of traversable wormholes is, for the moment, entirely speculative.

So, perhaps the most important taming operation may be for
the birth of the Universe, because, as summarized in
Sec.~\ref{sec:Intro},
the Big Bang is truly observable, even if only indirectly
(in addition to the detected CMBR and helium nuclei,
there is perhaps the possibility of observing also
a cosmological gravitational-wave background;
cf. Chap.~22 in Ref.~\refcite{Maggiore2018}).
For this reason, we have focussed on the tamed Big Bang in the
present review,
while referring to an earlier review~\cite{Klinkhamer2014-mpla}
for the tamed black hole singularity and to
\ref{app:Recap exotic-matter-wormhole}
and \ref{app:Defect-wormhole solution} here
for regular behavior at the wormhole throat
from either exotic matter or an exotic spacetime.

Special focus of this review
has been on the mathematics needed for
a proper description of the three-dimensional spacetime defect
(a submanifold with
vanishing  determinant of the metric $g_{\mu\nu}$).
In our interpretation, the spacetime points of the defect
differ essentially from the other spacetime points.
In fact, we are inspired by the well-known physics of
crystallographic defects in atomic crystals, as mentioned in
the second paragraph of Sec.~\ref{sec:Nature-and-origin}.

In order to understand the nature of the three-dimensional
cosmological defect (assuming its relevance for the
actual Universe), it may be crucial to know more about
its origin, most likely from a quantum-gravity phase
in the very early Universe.
Regrettably, we know nothing for sure about such a
quantum-gravity phase.
In Sec.~\ref{sec:Nature-and-origin},
we have presented a heuristic idea and sketched
a toy-model calculation within the framework of nonperturbative
superstring theory in the guise of a particular matrix model.

%%%%\newpage%%tmp
However,  even with 
the ultimate origin of a Big Bang spacetime defect shrouded in mystery, 
it might still be interesting to explore
the characteristics of the suggested defect
cosmology~\cite{Klinkhamer2019-prd,Klinkhamer2020-more,%
KlinkhamerWang2019-cosm,KlinkhamerWang2020-pert,Wang2021,Battista2021}
(many other types of bouncing cosmology have been discussed
in the literature;
see, e.g., Refs.~\refcite{IjjasSteinhardt2018,Brandenberger2023}
with further references therein).

Defect cosmology can,  in fact, give
a more less realistic description of the expansion
history of \emph{our} Universe with a ``regularized'' Big Bang at
early times ($t=0$), a matter-dominated phase at intermediate
times ($0< t \lesssim 10^{10}\,\text{yr}$),
and a vacuum-dominated phase with
accelerated expansion at late times ($t \gtrsim 10^{10}\,\text{yr}$).
For this particular description, we use the new metric
\textit{Ansatz}~\eqref{eq:RWK-ds2} 
and a homogeneous multi-component perfect fluid,   
namely a mix of
%various energy density and pressure components:
relativistic matter
with $\rho_\text{M,rel} = 3\,P_\text{M,rel} >0$,
nonrelativistic matter
with $\rho_\text{M,nonrel}>0$ and $P_\text{M,nonrel} = 0$,
and a ``vacuum'' contribution with
$\rho_\text{V} = -P_\text{V} = \Lambda >0$ as discussed in
the last paragraph of Sec.~\ref{subsec:Metric and Einstein field equation}.
It is also possible to add a
\emph{dynamic} vacuum component, possibly from a
special type of scalar field (for example, the $q$ field
discussed in Ref.~\refcite{KlinkhamerVolovik2008a}
and subsequent papers), without changing the defect
behavior at $t=0$  (see also the discussion in App.~B
of Ref.~\refcite{Klinkhamer2019-prd} for the case of a
constant vacuum component).

%%%%%%%%%%%%%%%%%%\newpage%%tmp
Yet, the most important prediction of defect cosmology
would be 
that there exists \emph{another} world,  
with or without parity reversal
(as discussed in the last paragraph of
Sec.~\ref{subsec:Defect cosmological tetrads and spin connection}).
\emph{Both} worlds, the two branches of the full cosmological solution,
can be considered to be \emph{expanding}
in the direction of the thermodynamic arrow of time, defined
by growing matter density perturbations and increasing
entropy~\cite{BoyleFinnTurok2018,KlinkhamerWang2020-pert}
(here, briefly recalled in
Sec.~\ref{subsec:Coordinate time and thermodynamic time}).
If this is indeed the correct physical interpretation,
then it appears that direct (classical) communication
between the two worlds is impossible, leaving only quantum effects,
as will be explained in
Sec.~\ref{subsec:Direct communication or not}.
(The specific results of Sec.~\ref{sec:Communication-between-the-two-worlds} 
may also be relevant to other cosmological-bounce models
with different explanations of the effective
energy-condition violations at the bounce.)
Given the two branches of defect cosmology and
barring the possibility of direct communication between them,
there may still be \emph{other} implications
than quantum entanglement and it
makes sense  %%is certainly worthwhile
to search for possible new observables.

\appendix

%%\newpage%%tmp
\section{Recap: Exotic-matter wormhole}
\label{app:Recap exotic-matter-wormhole}

Morris and Thorne (MT) have shown~\cite{MorrisThorne1988} that 
traversable wormholes
could, in principle, exist if exotic matter could be provided for.
They also gave a relatively simple metric for
such a traversable wormhole:
\begin{eqnarray}
\label{eq:exotic-WH}
&&ds^{2}\,\Big|^\text{(EBMT)}
\equiv
g_{\mu\nu}(x)\, dx^\mu\,dx^\nu \,\Big|^\text{(EBMT)}
\nonumber\\[1mm]
&& =
- dt^{2} + dl^{2}
+ \left(b_{0}^{2} + l^{2}\right)\,
  \Big[ d\theta^{2} + \sin^{2}\theta\, d\phi^{2} \Big]\,,
\end{eqnarray}
with a positive length scale $b_{0}$.
The dimensionless angular coordinates
$\theta \in [0,\,\pi]$ and $\phi \in [0,\,2\pi)$
are the standard spherical %%polar
coordinates, while the dimensional coordinates $t$ and $l$
range over $(-\infty,\,\infty)$. The same metric has also
been considered in two earlier papers~\cite{Ellis1973,Bronnikov1973},
which explains the first two letters `EB' in the superscript  
of \eqref{eq:exotic-WH}.

%%FRK MEMO: need to flip Riemann and Ricci signs in my WH paper,
%%in order to get Einstein equation}
%%R_{\mu\nu} - \half\, g_{\mu\nu}\,R = MINUS 8\pi G\:T_{\mu\nu}\,,

\begin{figure}[t]
\vspace*{0mm}   %%renamed FIG04-v4.eps
\centerline{\includegraphics[width=0.75\textwidth]{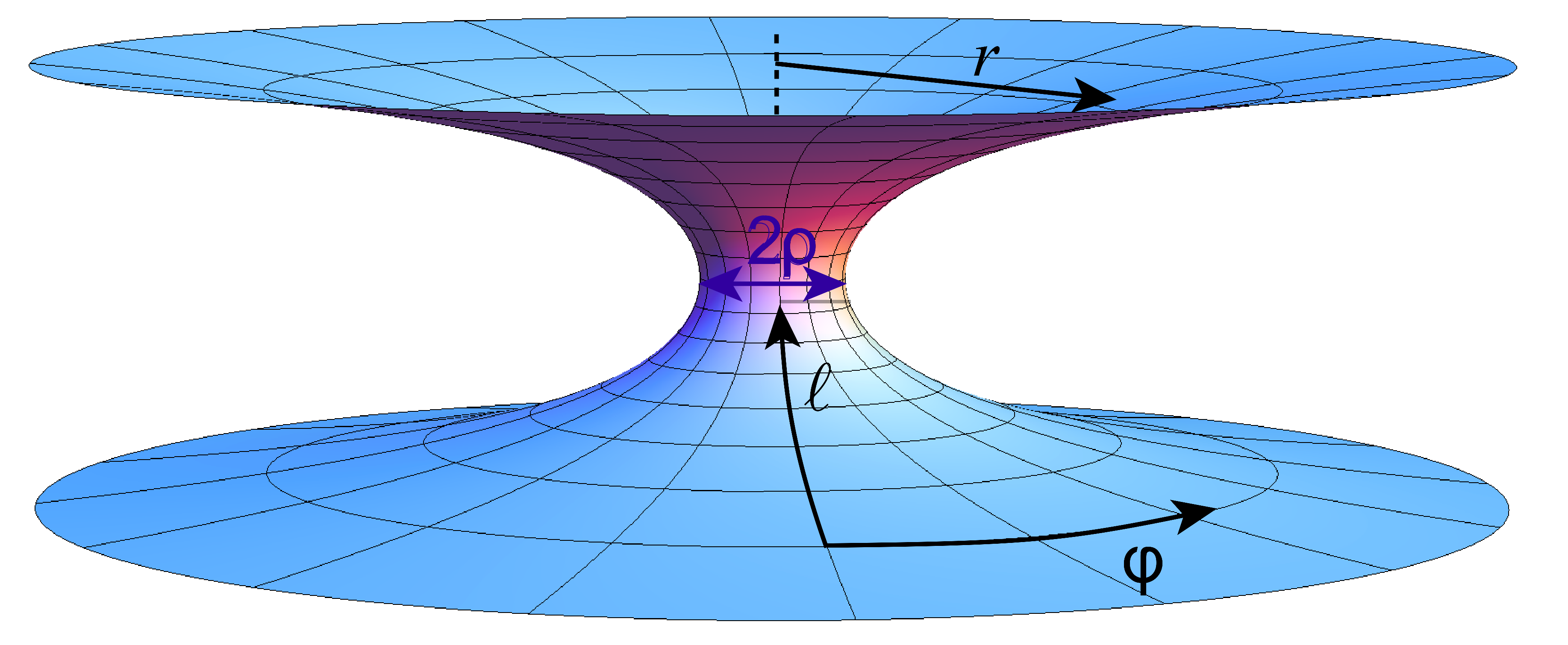}}
%%proceedings-basic2023-klinkhamer-Fig2-v1.eps
%%{Oliver_James-1502.03809v3-Fig01.eps}
\vspace*{8pt}
\caption{Embedding diagram of a wormhole spacetime with metric
\eqref{eq:exotic-WH} for constant values of the time coordinate $t$
and in the equatorial slice ($\theta=\pi/2$).
The notation used in the figure differs somewhat from the one in the text,
specifically $2 \rho \equiv 2 b_{0}>0$
and $r \equiv \sqrt{b_{0}^{2} + l^{2}}\,$.
[Image credit: Ref.~\protect\refcite{James-etal2015}]}  
\protect\label{fig:Ellis-wormhole}
\vspace*{-0mm}
\end{figure}

The resulting Ricci and Kretschmann curvature scalars are
\begin{subequations}\label{eq:exotic-mat-WH-R-K}
\begin{eqnarray}
R\,\Big|^\text{(EBMT)}
&=&
%%-  %% flipped minus sign
2\;\frac{b_{0}^{2}}
         {\left(b_{0}^{2} + l^{2}\right)^{2}}\,,
\\[0mm]
K\,\Big|^\text{(EBMT)} &=&
%% sign unchanged
12\;\frac{\left(b_{0}^{2}\right)^{2}}
         {\left(b_{0}^{2} + l^{2}\right)^{4}}\,,
\end{eqnarray}
\end{subequations}
both of which are finite throughout and
vanishing as $l \to \pm\infty$.
Indeed, two distinct flat Minkowski spacetimes are
approached for $l \to \pm\infty$,
with the wormhole throat at $l=0$ connecting them;
see Fig.~\ref{fig:Ellis-wormhole}.
Moreover, this wormhole can be shown to be traversable:
an intrepid human traveller can survive the trip
from positive values of $l$ to negative values, or \textit{vice versa},
provided the length scale $b_{0}$ is large enough
(see also Sec.~III~F~4 in Ref.~\refcite{MorrisThorne1988}).

%%%%\newpage%%tmp
But can this particular wormhole metric
arise as a solution of the Einstein equation?
Morris and Thorne suggested to explore what type
of matter would be needed.
With the metric (\ref{eq:exotic-WH})
for a traversable wormhole,
the Einstein equation \eqref{eq:Einstein-eq}
requires the following
components of the energy-momentum tensor~\cite{MorrisThorne1988}:
\begin{subequations}\label{eq:T-components-exotic-mat-WH}
\begin{eqnarray}
\label{eq:T-tt-component-exotic-mat-WH}
T^{\,t}_{\;\;\,t}\,\Big|^\text{(EBMT)}
&=& \frac{1}{8\pi G}\;
\frac{b_{0}^{2}}{\left(b_{0}^{2} + l^{2}\right)^{2}}\,,
\\[1mm]
T^{\,l}_{\;\;\,l}\,\Big|^\text{(EBMT)}
&=& -\frac{1}{8\pi G}\;
\frac{b_{0}^{2}}{\left(b_{0}^{2} + l^{2}\right)^{2}}\,,
\\[1mm]
T^{\,\theta}_{\;\;\,\theta}\,\Big|^\text{(EBMT)}
&=& \frac{1}{8\pi G}\;
\frac{b_{0}^{2}}{\left(b_{0}^{2} + l^{2}\right)^{2}}\,,
\\[1mm]
T^{\,\phi }_{\;\;\,\phi}\,\Big|^\text{(EBMT)}
&=& \frac{1}{8\pi G}\;
\frac{b_{0}^{2}}{\left(b_{0}^{2} + l^{2}\right)^{2}}\,,
\end{eqnarray}
\end{subequations}
with all other components vanishing.

These  energy-momentum components can be seen to
correspond to some form of exotic matter.
The energy density, for example, is given
by $\rho = T^{\,tt} =-T^{\,t}_{\;\;\,t}$ and
we have $\rho< 0$ from (\ref{eq:T-tt-component-exotic-mat-WH}),
which is rather unusual.
Indeed, the radial null vector $\overline{k}^{\,\mu}=(1,\, 1,\,  0,\,0)$
gives the following inequality:
\begin{equation}
\label{eq:exotic-mat-WH-NEC-radial}
T^{\,\mu}_{\;\;\,\nu}\,
\overline{k}_{\mu}\,\overline{k}^{\,\nu}\,\Big|^\text{(EBMT)}
= \frac{1}{8\pi G}\;
\frac{b_{0}^{2}}{\left(b_{0}^{2} + l^{2}\right)^{2}}\;\big[-1-1\big] < 0\,,
\end{equation}
which corresponds to a violation of the Null-Energy-Condition (NEC),
also mentioned in the last sentence of
Sec.~\ref{subsec:RW metric and Friedmann equations}
for the cosmological context.
This exotic matter keeps the wormhole throat at $l=0$ open:
without exotic matter,
the wormhole throat would collapse, which is what happens for the
Einstein--Rosen bridge~\cite{EinsteinRosen1935}.
The crucial question is whether or not the needed exotic matter really
exists in Nature (see, e.g., Ref.~\refcite{Visser1996} for an extensive discussion).

%%\newpage%%tmp
\section{Defect-wormhole solution}
\label{app:Defect-wormhole solution}

\subsection{Defect-wormhole metric}
\label{subapp:Defect wormhole  metric}

Starting form the
exotic-matter wormhole metric \eqref{eq:exotic-WH},
we propose a new wormhole metric~\cite{Klinkhamer2023a}:
\begin{eqnarray}
\label{eq:defect-WH}
&&ds^{2}\,\Big|^\text{(EBMTK)}
\equiv
g_{\mu\nu}(x)\, dx^\mu\,dx^\nu \,\Big|^\text{(EBMTK)}
\nonumber\\[1mm]
&&=
- dt^{2} + \frac{\xi^{2}}{\xi^{2}+\lambda^{2}}\;d\xi^{2}
+ \left(b_{0}^{2} + \xi^{2}\right)\,
  \Big[ d\theta^{2} + \sin^{2}\theta\, d\phi^{2} \Big]\,,
\end{eqnarray}
with positive length scales $b_{0}$ and $\lambda$. The
coordinates $t$ and $\xi$ range over $(-\infty,\,\infty)$,
while $\theta$ and $\phi$ are the standard spherical polar
coordinates.

The Ricci and Kretschmann curvature scalars of the
new wormhole metric are
\begin{subequations}\label{eq:defect-WH-R-K}
\begin{eqnarray}
R\,\Big|^\text{(EBMTK)} &=&
%%-  %% flipped minus sign
2\;\frac{b_{0}^{2}-\lambda^{2}}
         {\left(b_{0}^{2} + \xi^{2}\right)^{2}}\,,
\\[1mm]
K\,\Big|^\text{(EBMTK)} &=&
12\;\frac{\left(b_{0}^{2}-\lambda^{2}\right)^{2}}
         {\left(b_{0}^{2} + \xi^{2}\right)^{4}}\,,
\end{eqnarray}
\end{subequations}
which are smooth and finite throughout. These curvature scalars
drop to zero for $\xi \to \pm\infty$.

The metric $g_{\mu\nu}(x)$ from (\ref{eq:defect-WH})
is \emph{degenerate} with a vanishing determinant
$g(x) \equiv \det[g_{\mu\nu}(x)]$ at $\xi=0$
(the coordinate singularities at $\theta=0,\,\pi$
can be removed by appropriate coordinate transformations,
as detailed in the second paragraph of Sec.~3.1 of Ref.~\refcite{Klinkhamer2023a}).
In physical terms, this three-dimensional hypersurface at $\xi=0$
corresponds to a ``spacetime defect,''
similar to the cosmological spacetime defect discussed in
Sec.~\ref{subsec:Defect cosmological metric and modified Friedmann equations}
and the black hole spacetime defect reviewed in Ref.~\refcite{Klinkhamer2014-mpla}.

%%%%%%\newpage%%tmp
In the spirit of Morris and Thorne's engineering
approach~\cite{MorrisThorne1988},
the Einstein equation \eqref{eq:Einstein-eq}
for this new metric \eqref{eq:defect-WH} then requires
the following nonzero energy-momentum-tensor
components~\cite{Klinkhamer2023a}:
\begin{subequations}\label{eq:T-UPDOWNcomponents-degenmetric-special}
\begin{eqnarray}
\label{eq:T-UPtDOWNt-component-degenmetric-special}
T^{\,t}_{\;\;\,t}\,\Big|^\text{(EBMTK)}
&=& \frac{1}{8\pi G}\;
\frac{b_{0}^{2}-\lambda^{2}}{\left(b_{0}^{2}+\xi^{2}\right)^{2}}\,,
\\[1mm]
T^{\,\xi}_{\;\;\,\xi}\,\Big|^\text{(EBMTK)}
&=& -\frac{1}{8\pi G}\;
\frac{b_{0}^{2}-\lambda^{2}}{\left(b_{0}^{2}+\xi^{2}\right)^{2}}\,,
\\[1mm]
T^{\,\theta}_{\;\;\,\theta}\,\Big|^\text{(EBMTK)}
&=& \frac{1}{8\pi G}\;
\frac{b_{0}^{2}-\lambda^{2}}{\left(b_{0}^{2}+\xi^{2}\right)^{2}}\,,
\\[1mm]
T^{\,\phi }_{\;\;\,\phi}\,\Big|^\text{(EBMTK)}
&=& \frac{1}{8\pi G}\;
\frac{b_{0}^{2}-\lambda^{2}}{\left(b_{0}^{2}+\xi^{2}\right)^{2}}\,.
\end{eqnarray}
\end{subequations}
%%%%\newpage%%tmp
\noindent An important point to mention is that
the Einstein equation \eqref{eq:Einstein-eq}
for the new metric \eqref{eq:defect-WH} at $\xi=0$ is to be defined
by continuous extension from its limit $\xi \to 0$.
This procedure is the same as discussed in
Sec.~\ref{subsec:Mathematics-Continuous-extension}
for the cosmological defect. For the defect wormhole, the procedure will
be elaborated upon in \ref{subapp:Defect-WH-behavior-at-throat}.

The basic structure of the energy-momentum-tensor components
(\ref{eq:T-UPDOWNcomponents-degenmetric-special})
is identical to that of
\eqref{eq:T-components-exotic-mat-WH},
but the crucial difference is that the positive
factor $b_{0}^{2}$ in the numerators
of the previous results has been replaced by a
factor $(b_{0}^{2}-\lambda^{2})$, which can have either sign.
Indeed, as $\lambda^{2}$ increases above $b_{0}^{2}$,
we no longer require the presence of exotic matter.
For example, we have from (\ref{eq:T-UPtDOWNt-component-degenmetric-special})
that $\rho =-T^{\,t}_{\;\;\,t} > 0$ for $\lambda^{2} > b_{0}^{2}\,$.
More generally, for an arbitrary null vector $k^{\,\mu}$
and parameters $\lambda^{2} \geq b_{0}^{2}\,$,
we can readily establish the inequality
\begin{equation}
\label{eq:exotic-mat-WH-NEC}
T^{\,\mu}_{\;\;\,\nu}\,k_{\mu}\,k^{\,\nu}\,
\Big|^\text{(EBMTK)}_{\lambda^{2} \geq b_{0}^{2}}
\geq 0\,,
\end{equation}
and the NEC is satisfied, without need of exotic matter.

Even the case of no matter at all is covered by
the new metric (\ref{eq:defect-WH}) for $\lambda^{2}=b_{0}^{2}$,
which has a vanishing energy-momentum tensor,
\begin{equation}
\label{eq:T-UPmuUPnu-component-degenmetric-special}
T^{\,\mu}_{\;\;\,\nu}\,\Big|^\text{(EBMTK)}_{\lambda^{2}=b_{0}^{2}}
= 0\,,
\end{equation}
and vanishing  curvature scalars according to (\ref{eq:defect-WH-R-K}).
This defect spacetime is flat
but it has a different topology than Minkowski spacetime,
specifically it has a different spatial orientability
(cf. Sec.~2.2 of Ref.~\refcite{Klinkhamer2023b}).

Further discussion
of  the new wormhole metric \eqref{eq:defect-WH}
is given in the follow-up papers~\cite{Klinkhamer2023b,%
Wang2023,Ahmed2023-JCAP,Ahmed2023-APPB,%
Feng2023,BainesGaurVisser2023}
(the criticism of the last two papers will be
addressed in \ref{subapp:Defect-WH-behavior-at-throat}).
A short introduction to the basic idea of the defect-wormhole
is given in the review~\cite{Klinkhamer2023-BJP}.
Let us make a final comment on the issue of
traversability for the defect wormhole with
appropriate normal matter ($\lambda^{2} > b_{0}^{2}$)
or without matter at all ($\lambda^{2} = b_{0}^{2}$).
Safe passage of a human traveller certainly
requires a large enough value of the
length scale $b_{0}$ (even for the vacuum case,
the traveller should comfortably fit in and pass through,
which requires $b_{0} \gg 1\,\text{m}$).
%, while tidal forces can be reduced further by having small
%enough $\lambda^{2} - b_{0}^{2} \geq$

%%%%\newpage%%tmp
\subsection{Defect-wormhole tetrads and spin connection}
\label{subapp:Defect-wormhole tetrads and spin connection}

For a closer analysis, we now give the tetrad $e^{a}_{\mu}(x)$
and spin connection $\omega^{\phantom{z}a}_{\mu\phantom{z}b}(x)$
corresponding to the defect-wormhole metric (\ref{eq:defect-WH}),
reproducing previous results~\cite{Klinkhamer2023b} for the
vacuum case $\lambda^{2}=b_{0}^{2}$.
Again, we will use the differential-form formalism in the notation
of Ref.~\refcite{EguchiGilkeyHanson1980}, as summarized in
Sec.~\ref{subsec:Tetrads, spin connection, and first-order vacuum equations}.

As regards the defect wormhole, we make the following
\textit{Ans\"{a}tze} for the dual basis (defined
by $e^{a} \equiv e^{a}_{\phantom{z}\mu}\,\text{d}x^{\mu}$
in terms of the tetrad $e^{a}_{\phantom{z}\mu}$)\,:
\begin{subequations}\label{eq:defect-WH-tetrad}
\begin{eqnarray}
\label{eq:defect-WH-tetrad-a-is-0}
e^{0}\,\Big|^\text{(EBMTK)}
&=& \text{d}t\,,
\\[1mm]
\label{eq:defect-WH-tetrad-a-is-1}
e^{1}\,\Big|^\text{(EBMTK)}
&=&  \frac{\xi}{\sqrt{\lambda^{2} + \xi^{2}}}\; \text{d}\xi\,,
\\[1mm]
e^{2}\,\Big|^\text{(EBMTK)}
&=&  \sqrt{b_{0}^{2} + \xi^{2}}\; \text{d}\theta\,,
\\[1mm]
e^{3}\,\Big|^\text{(EBMTK)}
&=& \sqrt{b_{0}^{2} + \xi^{2}}\;\sin\theta  \;  \text{d}\phi\,,
\end{eqnarray}
\end{subequations}
and for the nonzero components of the connection 1-form
(defined by $\omega^{a}_{\phantom{z}b}
\equiv \omega^{\phantom{z}a}_{\mu\phantom{z}b}\,\text{d}x^{\mu}$)\,:
\begin{eqnarray}
\label{eq:defect-WH-connection}
&&
\left\{
\omega^{2}_{\phantom{z}1},\,
\omega^{3}_{\phantom{z}1},\,
\omega^{3}_{\phantom{z}2}
\right\}\,\Big|^\text{(EBMTK)}=
\left\{
-\omega^{1}_{\phantom{z}2},\,
-\omega^{1}_{\phantom{z}3},\,
-\omega^{2}_{\phantom{z}3}
\right\}\,\Big|^\text{(EBMTK)}
\nonumber\\[1mm]
&&=
\left\{
\sqrt{\frac{\lambda^{2} + \xi^{2}}{b_{0}^{2} + \xi^{2}}}\,\text{d}\theta,\,
\sqrt{\frac{\lambda^{2} + \xi^{2}}{b_{0}^{2} + \xi^{2}}}\,\sin\theta  \;
\text{d}\phi,\,
\cos\theta  \;  \text{d}\phi
\right\}\,.
\end{eqnarray}
These \textit{Ans\"{a}tze} are, by construction,
consistent with the metricity and no-torsion conditions.
Furthermore, these tetrads and connections are perfectly smooth,
notably at the wormhole throat $\xi=0$.

The corresponding curvature 2-form has the following nonzero components:
\begin{subequations}\label{eq:defect-WH-curvature}
\begin{eqnarray}
&&
\left\{
R^{2}_{\phantom{z}1},\,
R^{3}_{\phantom{z}1},\,
R^{3}_{\phantom{z}2}
\right\}\,\Big|^\text{(EBMTK)}
=
\left\{
-R^{1}_{\phantom{z}2},\,
-R^{1}_{\phantom{z}3},\,
-R^{2}_{\phantom{z}3}
\right\}\,\Big|^\text{(EBMTK)}
\nonumber\\[1mm]
&&=
\left\{
-F^{\prime}
\,\text{d}\xi \wedge \text{d}\theta,\,
F^{\prime}\,
\sin\theta  \;  \text{d}\xi \wedge \text{d}\phi,\,
-\left(b_{0}^{2} - \lambda^{2}\right)\,
\sin\theta  \;  \text{d}\theta \wedge \text{d}\phi
\right\}\,,
\end{eqnarray}
with
\begin{eqnarray}
F^{\prime}
&\equiv&
\frac{\text{d}}{\text{d}\xi}\,
\sqrt{
\frac{\lambda^{2} + \xi^{2}}
     {b_{0}^{2} + \xi^{2}}
     }
=
\left(b_{0}^{2} - \lambda^{2}\right)\,
\frac{\xi}{(b_{0}^{2} + \xi^{2})^{2}}\;
\sqrt{
\frac{b_{0}^{2} + \xi^{2}}
     {\lambda^{2} + \xi^{2}}
     }\,.
\end{eqnarray}
\end{subequations}

The tetrad component $e^{1}_{\phantom{z}\mu}(x)$
from \eqref{eq:defect-WH-tetrad-a-is-1} has
a factor $\xi/(\lambda^{2} + \xi^{2})^{1/2}$, which
is essential for obtaining %%getting
a smooth solution.
A different factor $|\xi|/(\lambda^{2} + \xi^{2})^{1/2}$
would give, from the
no-torsion condition \eqref{eq:first-order-eqs-no-torsion},
discontinuous terms $\xi/|\xi|$ in the connection components
$\omega^{2}_{\phantom{z}1}$ and $\omega^{1}_{\phantom{z}2}$.
There would then result delta-function singularities in the
curvature 2-form components
$R^{2}_{\phantom{z}1} =-R^{1}_{\phantom{z}2} \propto
\delta(\xi)\,\text{d}\xi \wedge \text{d}\theta$.
Apparently, a smooth defect wormhole spacetime
requires a change in the spatial orientability
of the two asymptotically-flat spaces~\cite{Klinkhamer2023b}.

%%%%\newpage%%tmp
\subsection{Defect wormhole: Behavior at the wormhole throat}
\label{subapp:Defect-WH-behavior-at-throat}

The tetrad components \eqref{eq:defect-WH-tetrad} and
the spin connection components \eqref{eq:defect-WH-connection}
of the defect wormhole
are  perfectly smooth at the wormhole throat $\xi=0$ and the
same holds for the curvature 2-form
components \eqref{eq:defect-WH-curvature}.
For the vacuum defect wormhole with parameters $\lambda^{2}=b_{0}^{2}$,
this suffices as
the first-order formalism \eqref{eq:first-order-eqs} is equivalent to
the usual second-order formalism \eqref{eq:Einstein-eq}
with $T_{\mu\nu}^\text{\,(M)}=0$, as explained in, e.g.,
Refs.~\refcite{Wald1984,Horowitz1991}.
With a nonvanishing presence of normal matter
and metric parameters $\lambda^{2} > b_{0}^{2}$,
the first-order formalism is not directly applicable,
but the second-order formalism
with the continuous-extension procedure
also shows the regular behavior at the wormhole throat.

There have, however, been two recent
papers~\cite{Feng2023,BainesGaurVisser2023}, which are critical
about the vacuum defect wormhole.
It appears that these papers result from a misunderstanding,
as illustrated by the last sentence
in the Abstract of Ref.~\refcite{Feng2023}:
``I point out that the smoothness of the metric tensor components is
deceptive, and that in general relativity, such metrics must be sourced
by exotic thin shells.''
The decisive observation, now, is that
%But
we do \emph{not} use standard general relativity,
as clarified in Sec.~\ref{subsec:Mathematics-Continuous-extension}
and the first paragraph of Sec.~\ref{sec:Nature-and-origin}.
This point has also been mentioned in some of our previous papers, with
perhaps the clearest statement in
the paragraph ``Observe that $\ldots$'' starting under Eq.~(2.29) of
Ref.~\refcite{KlinkhamerSorba2014}.
In short, the crucial inputs for the degenerate metrics considered
here are the $C^\infty$ metric $g_{\mu\nu}$
and the continuous-extension procedure.

It is also stated in Ref.~\refcite{Feng2023} that
``The Einstein tensor acquires a nonvanishing term proportional to
$\delta(\rho)$, and $\ldots$'', where $\rho$ replaces our coordinate
$\xi$ from \eqref{eq:defect-WH};
in the following discussion, we will stick to our notation $\xi$.
Now, the calculation for the vacuum defect wormhole
in \ref{subapp:Defect wormhole  metric}
has shown that the Einstein tensor $G_{\mu\nu}$
vanishes at $\xi=0$ by continuous extension.
Moreover, the spin connection components \eqref{eq:defect-WH-connection}
are perfectly smooth at $\xi=0$ and
the curvature 2-form components \eqref{eq:defect-WH-curvature}
vanish identically for the vacuum case
with parameters $\lambda^{2}=b_{0}^{2}$.

Perhaps the delta-function from the last quote of
Ref.~\refcite{{Feng2023}} and those mentioned in
Ref.~\refcite{BainesGaurVisser2023} result from
the fact that these authors
consider directly $R_{\mu\nu}$ at $\xi=0$,
whereas we implicitly work with
$g^2\,R_{\mu\nu}=W_{\mu\nu}$
%or, more explicitly, with $W_{\mu\nu}$
and the continuous-extension procedure $\xi\to 0$;
see Refs.~\refcite{EinsteinRosen1935,Horowitz1991,Guenther2017}
for details and our brief summary
in Sec.~\ref{subsec:Mathematics-Continuous-extension}.
According to \eqref{eq:Horowitz1991-Eq.4.3},
directly working with $R_{\mu\nu}$  at $\xi=0$ gives $0/0$
and it is then no surprise that
a nonvanishing (or even infinite) result has been found in
the papers~\cite{Feng2023,BainesGaurVisser2023}.
Incidentally, our calculations have also shown potential
delta-function contributions to the curvature 2-form
if the ``wrong'' tetrad is chosen, whereas there are no
such contributions for the ``right'' tetrad;
see the last paragraph of
\ref{subapp:Defect-wormhole tetrads and spin connection}
for the wormhole case and the
paragraph containing \eqref{eq:RWK-curvature-m-0-BAD}
for the cosmological case.

%%%%\newpage%%tmp
The issue is really not mathematical but physical, namely
what physical situation is to be described.
As we have stated in the first paragraph of
Sec.~\ref{sec:Nature-and-origin},
``the spacetime points of the defect submanifold must be treated
differently than all other spacetime points,''
which is the physical basis for our appeal
to the mathematical procedure of continuous extension.
Furthermore, if we consider the vacuum case, then there is no
matter whatsoever to make a thin wall from
(actually, Einstein and Rosen~\cite{EinsteinRosen1935} mention
a thin wall of \emph{fictitious} matter) and
we have to stick with the vacuum-defect-wormhole metric as is,
that is with a three-dimensional spacetime defect and the corresponding
discontinuities (e.g., in the extrinsic curvature
and geodesic expansion; cf. Refs.~\refcite{Wang2021,Battista2021}).
Note also that many ``accepted theorems'' (in the terminology of
Ref.~\refcite{Feng2023}) do \emph{not} hold for the case of
degenerate metrics, as discussed in, for example,
Sec.~1 of Ref.~\refcite{Horowitz1991}
and Chapter~3 of Ref.~\refcite{Guenther2017}.

\section*{Acknowledgments}

It is a pleasure to thank my students and collaborators
over the last years and, in particular, Z.L.~Wang and E.~Battista
for discussions on defect cosmology 
and comments on the manuscript.  
The referee is thanked for a useful suggestion.

%%\newpage%%tmp

\end{document}